\documentclass[12pt]{article}
\pdfoutput=1

\usepackage{graphicx,amssymb,amsmath,empheq}
\usepackage{amsthm,authblk}
\usepackage{empheq}
\usepackage{comment}
\usepackage{subfig} 
\usepackage[title]{appendix}
\usepackage{hyperref}
\hypersetup{colorlinks = true, linkcolor=black, citecolor=black, urlcolor=blue}
\usepackage{url}

\theoremstyle{plain}

\usepackage{tikz,fp}
\usepackage{tikz-cd}
\usetikzlibrary{arrows}
\usetikzlibrary{intersections}
\usetikzlibrary{shapes.geometric}
\usetikzlibrary{decorations.pathmorphing, patterns,shapes,fixedpointarithmetic}
\usetikzlibrary{decorations.markings}

\pgfdeclarepatternformonly{south west lines}{\pgfqpoint{-0pt}{-0pt}}{\pgfqpoint{3pt}{3pt}}{\pgfqpoint{3pt}{3pt}}{
        \pgfsetlinewidth{0.4pt}
        \pgfpathmoveto{\pgfqpoint{0pt}{0pt}}
        \pgfpathlineto{\pgfqpoint{3pt}{3pt}}
        \pgfpathmoveto{\pgfqpoint{2.8pt}{-.2pt}}
        \pgfpathlineto{\pgfqpoint{3.2pt}{.2pt}}
        \pgfpathmoveto{\pgfqpoint{-.2pt}{2.8pt}}
        \pgfpathlineto{\pgfqpoint{.2pt}{3.2pt}}
        \pgfusepath{stroke}}

\tikzset{
  mid arrow/.style={postaction={decorate,decoration={
        markings,
        mark=at position .575 with {\arrow{stealth}}
      }}},
  near arrow/.style={postaction={decorate,decoration={
        markings,
        mark=at position .275 with {\arrow{stealth}}
      }}},
  far arrow/.style={postaction={decorate,decoration={
        markings,
        mark=at position .800 with {\arrow{stealth}}
      }}},
  snake arrow/.style={fixed point arithmetic, decorate, decoration={snake,amplitude=2pt, segment length=11pt},postaction={decoration={markings,mark=at position 0.625 with {\arrow{stealth}}},decorate}},
}

\usepackage[nosort]{cite}

\newenvironment{defn}[1][Definition]{\begin{trivlist}
\item[\hskip \labelsep {\bfseries #1}]}{\end{trivlist}}

\renewcommand{\bar}{\overline}
\renewcommand{\tilde}{\widetilde}
\renewcommand{\hat}{\widehat}
\renewcommand{\leq}{\leqslant}
\renewcommand{\geq}{\geqslant}
\renewcommand{\Re}{\operatorname{Re}}
\renewcommand{\Im}{\operatorname{Im}}

\newcommand{\Tr}{\operatorname{Tr}}
\newcommand{\Tc}{\operatorname{{\bf T}_c}}

\newcommand{\const}{\mathrm{const}}
\newcommand{\VF}{\Upsilon}
\newcommand{\OTOC}{\operatorname{OTOC}}

\newcommand{\dkap}{\delta\kern-1.25pt\varkappa}
\newcommand{\tG}{\tilde{G}}
\newcommand{\tVF}{\tilde{\VF}}

\newcommand{\R}{{\mathrm{R}}}
\newcommand{\A}{{\mathrm{A}}}
\newcommand{\K}{{\mathrm{K}}}
\newcommand{\W}{{\mathrm{W}}}

\newcommand{\SL}{\operatorname{SL}}

\newcommand{\RR}{\mathbb{R}}

\newcommand{\kap}{\varkappa}
\newcommand{\eps}{\varepsilon}
\newcommand{\ph}{\varphi}

\newcommand{\calJ}{\mathcal{J}}

\newcommand{\calL}{\mathcal{L}}

\newcommand{\calN}{\mathcal{N}}

\newcommand{\calS}{\mathcal{S}}

\newcommand{\Sch}{\operatorname{Sch}}
\newcommand{\sgn}{\operatorname{sgn}}

\makeatletter

\newcommand*{\wideboxed}[1]{\setlength{\fboxsep}{1ex}%
  \fbox{\m@th$\displaystyle#1$}}
\makeatother

\title{An obstacle to sub-AdS holography for SYK-like models}

\author[1]{Pengfei Zhang}
\author[2,1]{Yingfei Gu}
\author[1]{Alexei Kitaev}
\affil[1]{\normalsize\it California Institute of Technology, Pasadena, CA 91125, USA}
\affil[2]{\normalsize\it Harvard University, Cambridge, MA 02138, USA}

\date{December 2, 2020}

\begin{document}

\maketitle

\begin{abstract}
We argue that ``stringy'' effects in a putative gravity-dual picture for SYK-like models are related to the branching time, a kinetic coefficient defined in terms of the retarded kernel. A bound on the branching time is established assuming that the leading diagrams are ladders with thin rungs. Thus, such models are unlikely candidates for sub-AdS holography. In the weak coupling limit, we derive a relation between the branching time, the Lyapunov exponent, and the quasiparticle lifetime using two different approximations.
\end{abstract}

\vspace{30pt}

\tableofcontents
\section{Introduction}

Out-of-time-order correlators (OTOCs) have attracted significant attention for two reasons. First, they describe quantum chaos in a way comparable to classical chaos (at least, for systems with a large parameter $N$, where the early-time OTOCs are characterized by a Lyapunov exponent). The second reason has to do with black holes and gravity. In this setting, OTOCs serve as a unique probe of high energy physics, namely, gravitational scattering between incoming and outgoing particles at the event horizon. The formal theory of such scattering was initiated by Dray and 't Hooft~\cite{DtH85} and further developed by 't Hooft~\cite{tH87,tH90, tH96}, but its physical meaning was hard to grasp because there seemed to be no observable effects. Only later it was realized that the relevant quantity is an OTOC and that it provides a basis for comparison between black holes and other systems. Maximal chaos, namely, the relation $\kap=2\pi T$ between the Lyapunov exponent and temperature, is a hallmark of gravity~\cite{Kit.BPS}, whereas small corrections to $\kap$ may be attributed to stringy effects~\cite{ShSt14} or some other form of nonlocality. (The term ``maximal chaos'' is due to the inequality $\kap\leq 2\pi T$, which holds under very general assumptions~\cite{MSS15}.)

The Sachdev-Ye-Kitaev model realizing maximal chaos~\cite{SaYe93,Kit.KITP.1} has contributed to the concept of quantum gravity as a very general phenomenon, something like ``quantum thermodynamics''. We would like to move toward a more concrete understanding of gravity. The SYK model is relatively simple but has a very rough bulk-dual picture, as compared with the celebrated duality between certain supersymmetric conformal theories and superstrings in anti-de Sitter spaces~\cite{Mal97}. The main goal of this paper is to figure what exactly is missing. An abstract answer is known: the SYK model lacks a gap in the conformal dimension spectrum, which is the standard condition for sub-AdS locality~\cite{HPPS09}. We will try to give a more intuitive explanation, which could help to dissect the problem and find a way around.

The rest of the introduction is devoted to the comparison between bulk-local and bulk-nonlocal effects in the context of the SYK model. In the main part of the paper, we show that the nonlocal effects cannot be diminished within a more general setting. This should provide useful guidance for the future search of holographic models.

In the discussion of holography for the SYK model, we will focus on the soft mode~\cite{Kit.KITP.2,MS16-remarks,kitaev2018soft}. In its basic form, the holographic correspondence is between the Schwarzian action $I_{\Sch} =-N\alpha_S J^{-1} \int_{0}^{1/T}\Sch\bigl(e^{i\ph(\tau)},\tau\bigr)\,d\tau$ on the boundary and Jackiw-Teitelboim (JT) gravity in the bulk~\cite{Jen16,MSY16,EMV16}. The latter has the following action:
\begin{equation}
I_{\mathrm{JT}}
=\frac{1}{4\pi}\int_{D} \bigl(-\Phi R+U(\Phi)\bigr)\sqrt{g}\,d^2x
-\frac{1}{2\pi}\int_{\partial D}\Phi K\sqrt{g_{\ph\ph}}\,d\ph,\qquad
U(\Phi)=-2\Phi,
\end{equation}
where the dilaton field $\Phi$ is normalized is such a way that its extremal value in an on-shell configuration equals the entropy. (The zero-temperature entropy, which can be represented by a topological term, is not included.) However, this correspondence is too simple because the Schwarzian action is already local. A more interesting form of holography would be one that provides a bulk-local description of some nonlocal boundary physics.

Let us mention another, more compelling reason to regard the JT gravity degenerate. It has to do with Dray-'t Hooft ``shock waves''. In the JT theory, a gravitational shock can be eliminated by an $\SL(2,\RR)$ coordinate transformation on one of the two half-spaces (say, on the future of the shock), such that the only remaining effect is a boundary shift. As a consequence, the boundary representation of the shock generator is local; for example, a shock at the past horizon is represented by $L_{-1}=e^{\kap t}\partial_t$. This operator is applied to one side of the thermofield double. In the low-temperature limit of the SYK model, the thermofield double is described by the Wightman function $G^\W$; its perturbation by the shock is given by the eigenfunction of the retarded kernel, $\Upsilon^\R=L_{-1}^{(1)}G^\W$, where $L_{-1}^{(1)}$ denotes the action of $L_{-1}$ on the first variable of $G^\W$~\cite{kitaev2018soft,gu2019relation}. The shock generator is more physical than the effective action because it is an on-shell object. We expect it to be well-defined (but not necessarily boundary-local) for all maximally chaotic systems. In fact, it is boundary-nonlocal for AdS black holes in $3+1$ dimensions, and it would be very desirable to find a concrete quantum model with this property.

Unlike with microscopic models, boundary nonlocality is easily achieved in the dilaton gravity setting. To break the degeneracy of the JT theory, let us perturb the dilaton potential with a quadratic term:
\begin{equation}\label{dil_pot}
U(\Phi)=-2\Phi-a\Phi^2.
\end{equation}
Its holographic dual is present in the SYK model as a correction to the Schwarzian action~\cite{kitaev2018soft}. We will now make use of some formulas from the cited paper to quantify the corresponding physical effects, which we call bulk-local. Then we will compare them with the correction to the Lyapunov exponent, which is a measure of bulk nonlocality. This may seem too technical for an introduction, but all cumbersome factors will magically cancel, resulting in a simple figure of merit.

The most important manifestation of the quadratic term is in thermodynamics. So let us consider static, i.e.\ rotationally symmetric, on-shell field configurations. Recall that the dilaton potential $\Phi$ at the center is equal to the entropy (up to a constant term). Furthermore, $-U(\Phi)$ is the temperature (up to an arbitrary factor)~\cite{kitaev2018soft,Mal16}. Thus, the entropy as a function of temperature is
\begin{equation}\label{S_dilaton}
S=\const + \frac{1}{2}(cT) - \frac{a}{8}(cT)^2 + O(T^3),
\end{equation}
where $c$ is arbitrary. On the SYK side, we have the following expansion in powers of $N$, where the extensive part is further expanded in $T/J$~\cite{randmat,JeSu16,kitaev2018soft}:
\begin{equation}
\ln Z + E_0/T = N\biggl(\calS
+ 2\pi^2\alpha_S(T/J) - \frac{\pi^2}{6}\gamma(T/J)^2 + \cdots \biggr)+O(N^0).
\end{equation}
Here $E_0$ is the ground state energy, $\calS$ is the zero-temperature entropy per Majorana site, $\alpha_S$ is the coefficient in the Schwarzian action, and $\gamma$ is the coefficient in the correction to it. Neglecting the $O(N^{0})$ term, we obtain the entropy in the thermodynamic limit:
\begin{equation}\label{S_SYK}
S = \frac{\partial(T\ln Z)}{\partial T} = N\biggl(\calS
+ 4\pi^2\alpha_S(T/J) - \frac{\pi^2}{2}\gamma(T/J)^2 + \cdots \biggr).
\end{equation}
The comparison between equations \eqref{S_dilaton} and \eqref{S_SYK} gives the value of parameter $a$ in the dilaton potential:
\begin{equation}
a=\frac{\gamma}{16\pi^2\alpha_S^2N}
=\frac{9q^3}{-k'_c(2)\,\pi(q-1)(q-2)\tan(\pi/q)\,N},
\end{equation}
where we have used the expressions for $\alpha_S$ and $\gamma$ from table~1 in Ref.~\cite{kitaev2018soft}. The notation $k_c(h)$ stands for the eigenvalue of the conformal kernel; $k'_c(2)$ is its derivative at $h=2$.

We are now in a position to determine the relative strengths of bulk-local and bulk-nonlocal corrections to the JT theory. To characterize the former, we consider the two terms in the specific heat:
\begin{equation}\label{C_SYK}
C = T\frac{\partial S}{\partial T}
=N\bigl(4\pi^2\alpha_S(T/J) - \pi^2\gamma(T/J)^2 +\cdots\bigr).
\end{equation}
Denoting the second (relatively small) term by $-\delta C$, we get:
\begin{equation}
\frac{\delta C}{C}=\frac{\gamma}{4\alpha_S}(T/J)
=\frac{36\pi q^3\alpha_S}{-k'_c(2)\,\pi(q-1)(q-2)\tan(\pi/q)}\,(T/J).
\end{equation}
This should be compared with the finite-temperature correction to the Lyapunov exponent, $\kap=2\pi T-\delta\kap$. Taking the value of $\delta\kap$ from~\cite{MS16-remarks} or~\cite{gu2019relation}, we get:
\begin{equation}\label{f_merit}
\wideboxed{
\frac{\delta\kap/\kap}{\delta C/C}
=\frac{1}{3}\, \frac{-k'_{c}(2)}{k'_{R}(-1)}\,.
}
\end{equation}
This is the promised figure of merit. It would be interesting to know how general this equation is, beyond the SYK setting. In any case, a holographic model would be one for which the left-hand side of equation~\eqref{f_merit} is small.
For the SYK model, we have
\begin{equation}
-k_{c}'(2)-k_{R}'(-1)=\pi\tan(\pi/q)>0,
\end{equation}
and hence, $\frac{-k'_{c}(2)}{k'_{R}(-1)}>1$. Looking for holography among slightly more general models, one could try to either decrease $-k'_{c}(2)$ or increase $k'_{R}(-1)$. We will see that the second recipe does not work because
\begin{equation}
\frac{\kap}{2\pi T}\,k'_{R}\biggl(-\frac{\kap}{2\pi T}\biggr)\leq 2
\label{eqn: bound unit}
\end{equation}
for a large class of SYK-like models. The number $t_B=\frac{1}{2\pi T} k'_{R}\bigl(-\frac{\kap}{2\pi T}\bigr)$ is called ``branching time''~\cite{gu2019relation}; its definition and use do not require maximal chaos.

\subsection{Outline of the paper}

The remainder of this paper is organized as follows. In section~\ref{sec: pre}, we review the kinetic equation for OTOCs with general rung function and establish our conventions. In section~\ref{sec: branching}, we review the concept of branching time. In section~\ref{sec: bound}, we present the main result of this paper, the inequality \eqref{eqn: bound unit}. It is derived by relating the branching time to the winding speed of the phase of the Green function. The required upper bound for the winding speed is proved in appendix~\ref{appendix: proof}. In section~\ref{sec: weak}, we investigate the branching time for weakly coupled models and discuss its relation to the quasi-particle lifetime and the Lyapunov exponent.

\section{Preliminaries}
\label{sec: pre}

We adopt the notation of Ref.~\cite{gu2019relation}, namely, set $\beta=2\pi$ and denote the connected out-of-time-order correlator by $\OTOC(t_1,t_2,t_3,t_4)$. More explicitly, 
\begin{equation}
\label{OTOCdef}
\OTOC(t_1,t_2,t_3,t_4):=
\langle X_1(\theta_1) X_2(\theta_2) \rangle \langle X_3(\theta_3) X_4(\theta_4) \rangle \mp
\langle X_1(\theta_1) X_3(\theta_3) X_2 (\theta_2) X_4 (\theta_4) \rangle,
\end{equation}
where $\theta_j=\tau_j+it_j$ $(j=1,2,3,4)$ are complex variables with their real parts (i.e.\ imaginary times) fixed as follows:
\begin{equation}\label{imag time}
\theta_1 = it_1 +\pi\,,\quad \theta_2=it_2\,,\quad \theta_3=it_3+\frac{\pi}{2}\,,\quad \theta_4=it_4-\frac{\pi}{2}\,.
\end{equation}
The $-$ and $+$ signs in \eqref{OTOCdef} are for bosons and fermions respectively. Following \cite{kitaev2018soft}, we assume the single mode ansatz
\begin{equation}\label{otoc}
\OTOC(t_1,t_2,t_3,t_4) \approx \frac{e^{\varkappa (t_1+t_2-t_3-t_4)/2   }}{{C}} \VF^\R_{X_1,X_2}(t_{12}) \VF^\A _{X_3,X_4}(t_{34})\,,\qquad
t_{jk}:=t_j-t_k
\end{equation}
within the time window 
\begin{equation}\label{time window}
t_{\text{scr}}\gg\frac{t_1+t_2}{2}-\frac{t_3+t_4}{2}\gg\varkappa^{-1}\,, \quad t_{12} \sim t_{34} \sim 1\,,
\end{equation}
where the scrambling time $t_{\text{scr}}$ is the time scale at which non-linear effects appear. For example, in the low temperature limit of the SYK model, $C\sim \frac{N}{\beta J}$ and $t_{\text{scr}}\approx\beta\ln C$. Finally, $\VF^{\R}_{X,Y}(t)$ and $\VF^{\A}_{X,Y}(t)$ denote the retarded and advanced vertex functions, respectively.  

\begin{defn}[Conventions.]
In this paper, we use the condensed matter convention in defining Green functions; for example, the imaginary-time and the retarded/advanced Green functions for Majorana operators are defined as follows:
\begin{equation}
G(\tau,\tau')=-\langle\chi_j(\tau)\chi_j(\tau')\rangle\quad
\text{for }\tau>\tau', \qquad
G^{\R/\A}(t,t')=\mp i\theta(\pm(t-t'))\left\langle \{\chi_j(it),\chi_j(it')\}\right\rangle.
\end{equation}
Here the factors $-1$ and $\pm i$ are chosen such that the bare Green functions in the frequency domain\footnote{In this paper, we add an additional tilde for functions in the frequency domain.} have the standard form, $\tG_0(\omega)=\omega^{-1}$. Let us also remind the reader that $\tG^{\R}(\omega)$ and $\tG^{\A}(\omega)$ are complex conjugates of each other and that they can be analytically continued to the upper and lower complex half-planes, respectively. These analytic continuations have the property that $\tG^{\R}(z)=\tG^{\A}(z^*)^*$.
\end{defn}

\subsection{Rung function in SYK-like models}

Connected four-point functions, including OTOCs, are given by sums of ladder diagrams. The function $\OTOC(t_1,t_2,t_3,t_4)$ with  fixed $t_3$, $t_4$ satisfies a kinetic equation~\cite{Kit.KITP.1,MS16-remarks,MSW17,gu2019relation},
\begin{equation}\label{kinetic equation}
 \int dt_5\,dt_6\, K^\R (t_1,t_2,t_5,t_6)  \OTOC(t_5,t_6,t_3,t_4)
\approx \OTOC(t_1,t_2,t_3,t_4)\,,
\end{equation}  
with retarded kernel expressed as follows
\begin{equation}
K^\R(t_1,t_2,t_3,t_4) =  G^\R (t_{13})  G^\A (t_{42}) R(t_{34})
=\begin{tikzpicture}[baseline={([yshift=-4pt]current bounding box.center)}]
\draw[thick, mid arrow] (40pt,15pt)--(0pt,15pt);
\draw[thick, mid arrow] (0pt,-15pt)--(40pt,-15pt);
\draw[thick, snake arrow] (40pt,-15pt) -- (40pt,15pt);
\filldraw  (40pt,-15pt) circle (1pt) node[right]{\scriptsize $t_4$};
\filldraw  (40pt,15pt) circle (1pt) node[right]{\scriptsize $t_3$};
\filldraw  (0pt,-15pt) circle (1pt) node[left]{\scriptsize $t_2$};
\filldraw  (0pt,15pt) circle (1pt) node[left]{\scriptsize $t_1$};
\end{tikzpicture}\,.
\label{eqn: ladder kernel}
\end{equation}
Here $R$, represented by the vertical wavy line and called the \emph{rung function}, depends on the exact form of interactions in the model. In the most general case, a ``rung'' is a two-particle irreducible diagram that involves four times at its corners. However, we consider the class of models where $R$ depends only on two times (or rather, their difference $t_{34}=t_3-t_4$) and refer to this assumption as the \emph{thin rung approximation}. For example, in the SYK model, the rung function is a product of Wightman functions, multiplied by the second moment of the random coupling:
\begin{equation}
R(t_{34})=J^2 (q-1) |G^\W(t_{34})|^{q-2}\,,
\qquad 
\text{i.e.}\quad 
\begin{tikzpicture}[baseline={([yshift=-4pt]current bounding box.center)}]
\draw[thick, snake arrow] (40pt,-15pt) -- (40pt,15pt);
\filldraw  (40pt,-15pt) circle (1pt);
\filldraw  (40pt,15pt) circle (1pt);
\node at (48pt,15pt) {\scriptsize $t_3$};
\node at (48pt,-15pt) {\scriptsize $t_4$};
\end{tikzpicture}=\,\frac{1}{(q-2)!}\,\,
\begin{tikzpicture}[baseline={([yshift=-4pt]current bounding box.center)}]
\filldraw  (40pt,-15pt) circle (1pt);
\filldraw  (40pt,15pt) circle (1pt);
\draw[thick, mid arrow] (40pt,-15pt)..controls (47pt,-7pt) and (47pt,7pt)..(40pt,15pt);
\draw[thick, mid arrow] (40pt,-15pt)..controls (33pt,-7pt) and (33pt,7pt)..(40pt,15pt);
\draw[thick, dotted] (40pt,-15pt) -- (40pt,15pt);
\node at (48pt,15pt) {\scriptsize $t_3$};
\node at (48pt,-15pt) {\scriptsize $t_4$};
\end{tikzpicture} 
\qquad \text{(in SYK)}\,.
\end{equation}
(The dotted line, representing the averaging over disorder, contributes the factor $(q-1)!J^2$.) With our choice of imaginary time shifts \eqref{imag time}, the Wightman function is related to the imaginary-time Green function as $G^\W(t)=G(it+\pi)$. In the generalized Keldysh formalism (see Appendix~\ref{appendix: kelydish}), the Wightman function is, essentially, the Keldysh Green function between two different contour folds (corresponding to the upper and lower rails in ladder diagrams), $G^\W= -\frac{i}{2} G^\K_{21}$. Meanwhile, the rung function can be expressed as the variational derivative of the Keldysh self-energy with respect to the Keldysh Green function, namely, $R = \frac{\delta \Sigma^\K_{21}}{\delta G^\K_{21}}$. 

One more thing the rung function is relevant to is the definition of the inner product of vertex functions,
\begin{equation}
\left( \VF^\A,\VF^\R \right) := \int \VF^{\R} (t) R(t) \VF^{\A}(t)\, dt\,,
\label{eqn: inner prod}
\end{equation}
which has been used in the ladder identity\cite{gu2019relation}.



\subsection{Kinetic equation in the frequency domain}

Following \cite{gu2019relation}, we define a variant of the kernel utilizing the time translation symmetry:
 \begin{equation}
 K_{\alpha}^{\R} (t,t') = \int K^{\R} \biggl( s+ \frac{t}{2}, s-\frac{t}{2},\frac{t'}{2}, -\frac{t'}{2}\biggr) e^{\alpha s} ds \,, \quad \alpha<0\,.
\label{eqn: variant kernel}
\end{equation}
We denote its largest eigenvalue by $k_{\R}(\alpha)$ and the corresponding eigenvector by $\VF_{\alpha}^{\R}(t)$, i.e. 
\begin{equation}
\int K_{\alpha}^{\R} (t,t')  \VF_{\alpha}^{\R}(t')\, dt' =k_{\R}(\alpha) \VF_{\alpha}^{\R}(t)\,.
\label{kernel_ev}
\end{equation}
In this notation, finding the Lyapunov exponent $\varkappa$ amounts to solving the equation $k_{\R}(-\varkappa)=1$. The retarded vertex function is given by the corresponding eigenvector $\VF^{\R}(t)=\VF_{-\varkappa}^{\R}(t)$.  Similarly, the advanced vertex function $\VF^\A=\VF^\A_{-\varkappa}$ is the corresponding eigenvector of the operator $K^\A_{\alpha}$ that is adjoint to $K^\R_{\alpha}$ with respect to the inner product \eqref{eqn: inner prod}.

For the purpose of this paper, it is useful to rewrite \eqref{kernel_ev} with $\alpha=-\varkappa$ in the frequency representation,
\begin{equation}
\int \tilde K_{-\varkappa}^{\R} (\omega,\omega') \tVF^{\R}(\omega') \frac{d\omega'}{2\pi} =\tVF^{\R} (\omega)\,,
\label{eqn: kinetic freq}
\end{equation}
where $\tVF^{\R} (\omega)$ is the Fourier transform of $\VF^{\R} (t)$, and $ \tilde K_{-\varkappa}^{\R} (\omega,\omega')= \int  K_{-\varkappa}^{\R}(t,t') e^{i(\omega t-\omega' t')} dt\, dt'$. Combining \eqref{eqn: ladder kernel} and \eqref{eqn: variant kernel}, we have the following explicit expression\footnote{where in the second step, we change the integration variables $(s, t, t')$ $\rightarrow$ $\big(s+\frac{t-t'}{2},-s+\frac{t-t'}{2},t'\big)$ with Jacobian $1$ and reorganize 
$
\omega t-\omega' t' + i\varkappa s$ as $\big(\omega+ i \frac{\varkappa}{2}\big) \big( s+ \frac{t-t'}{2} \big) + \big(\omega- i \frac{\varkappa}{2}\big) \big( -s+ \frac{t-t'}{2} \big) + (\omega-\omega')t'$.}
\begin{equation}
\begin{aligned}
\tilde K_{-\varkappa}^{\R} (\omega,\omega') &= \int  G^{\R} \Big( s+ \frac{t-t'}{2} \Big) G^{\A}\Big(  -s + \frac{t-t'}{2} \Big) R(t') e^{i(\omega t-\omega' t' + i \varkappa s)}\, ds\, dt\, dt' \\
& = \underbrace{\tG^{\R}\Big(\omega +  i \frac{\varkappa}{2}\Big) \tG^{\A}\Big(\omega - i \frac{\varkappa}{2}\Big)}_{=:W\left( \omega+i\frac{\varkappa}{2} \right) } \tilde R(\omega-\omega') \,.
\end{aligned}
\end{equation}
For later convenience, we define the $W$ function 
 as follows
\begin{equation}
\label{W_def}
\begin{tikzpicture}[baseline={([yshift=-4pt]current bounding box.center)}]
\draw[thick, mid arrow] (40pt,15pt)--(0pt,15pt);
\draw[thick, mid arrow] (0pt,-15pt)--(40pt,-15pt);
\filldraw  (40pt,-15pt) circle (1pt);
\filldraw  (40pt,15pt) circle (1pt);
\filldraw  (0pt,-15pt) circle (1pt) node[left]{\scriptsize $\tG^{\A}$};
\filldraw  (0pt,15pt) circle (1pt) node[left]{\scriptsize $\tG^{\R}$};
\node at (20pt,22pt) {\scriptsize $\omega+i \frac{\varkappa}{2}$};
\node at (20pt,-22pt) {\scriptsize $\omega-i \frac{\varkappa}{2}$};
\end{tikzpicture}
\qquad 
W\Bigl( \omega+i\frac{\varkappa}{2} \Bigr) = \tG^{\R}\Bigl(\omega +  i \frac{\varkappa}{2}\Bigr) \tG^{\A}\Bigl(\omega - i \frac{\varkappa}{2}\Bigr) = 
\left|  \tG^{\R}\Bigl(\omega +  i \frac{\varkappa}{2}\Bigr)  \right|^2 \geq 0
\end{equation}
where we have used 
the property $\tG^\R\big(\omega+i\frac{\varkappa(\mu)}{2}\big)^*=\tG^\A\big(\omega-i\frac{\varkappa(\mu)}{2}\big)$. Note that the equal sign in \eqref{W_def} is only possible on the real axis\footnote{because the imaginary part of the retarded Green function is always negative in the upper half-plane (c.f.\ Eq.~\eqref{eqn: imG}). On the other hand,  $G^\R$ can have zeros on the real axis.}, i.e.\ $W>0$ in the upper half-plane. 

In the Keldysh formalism, $W$ can be understood as the variation of the Keldysh Green function with respect to the Keldysh self-energy, $W=\frac{\delta G^\K_{21}}{\delta \Sigma^\K_{21}}$. In general, it depends on two variables, the center-of-mass frequency $i\kap$ and the relative frequency $\omega$. However, in the current problem, the center-of-mass frequency is set to be imaginary as we assume exponential growth, and the relative frequency is kept real. So we recast the two variables into a complex argument, $\omega+i\frac{\kap}{2}$.

Diagrammatically, the 
kinetic equation \eqref{eqn: kinetic freq}  can be represented as follows
\begin{equation}
\begin{tikzpicture}[baseline={([yshift=-4pt]current bounding box.center)}]
\draw[thick, mid arrow] (40pt,15pt)--(0pt,15pt);
\draw[thick, mid arrow] (0pt,-15pt)--(40pt,-15pt);
\draw[thick, snake arrow] (40pt,-15pt) -- (40pt,15pt);
\draw[thick, blue, mid arrow] (40pt,-15pt) .. controls (50pt,-10pt) and (50pt,10pt).. (40pt,15pt);
\filldraw  (40pt,-15pt) circle (1pt);
\filldraw  (40pt,15pt) circle (1pt);
\filldraw  (0pt,-15pt) circle (1pt) node[left]{\scriptsize $\tG^{\A}$};
\filldraw  (0pt,15pt) circle (1pt) node[left]{\scriptsize $\tG^{\R}$};
\node at (20pt,22pt) {\scriptsize $\omega+i \frac{\varkappa}{2}$};
\node at (20pt,-22pt) {\scriptsize $\omega-i \frac{\varkappa}{2}$};
\node at (55pt,0pt) {\scriptsize $\omega'$};
\node at (25pt,0pt) {\scriptsize $\omega-\omega'$};
\end{tikzpicture} = 
\begin{tikzpicture}[baseline={([yshift=-4pt]current bounding box.center)}]
\draw[thick,blue,mid arrow] (40pt,-15pt) .. controls (50pt,-10pt) and (50pt,10pt).. (40pt,15pt);
\filldraw  (40pt,-15pt) circle (1pt);
\filldraw  (40pt,15pt) circle (1pt);
\node at (55pt,0pt) {\scriptsize $\omega$};
\end{tikzpicture} \,.
\end{equation}
Here the blue curve represents the vertex function $\tVF^{\R}$, and  the frequency $\omega'$ in the loop should be integrated over.

To summarize, we have obtained the equation for the Lyapunov exponent $\kap$ and the retarded vertex function $\VF^\R$ in the frequency domain,
 \begin{equation}
 \label{eqn: key equation}
 \wideboxed{
W\Bigl( \omega+i\frac{\varkappa}{2} \Bigr) \int  \tilde R(\omega-\omega') \tVF^{\R}(\omega') \frac{d\omega'}{2\pi} =\tVF^{\R} (\omega) \,.
}
 \end{equation}

\subsection{Branching time and a generating function}
\label{sec: branching}

The branching time $t_B$ is defined in Ref.~\cite{gu2019relation} as  
\begin{equation}
t_B:=k_\R'(-\varkappa), 
\end{equation}
which measures the average rung separation. Physically, the branching time characterizes the sensitivity of the Lyapunov exponent to perturbations of the system. Let us imagine some theory with the retarded kernel $K^{\R}$, eigenvalue function $k_R(\alpha)$, and Lyapunov exponent $\varkappa$. Now we perturb the theory, which changes the retarded kernel, $K^{\R} \rightarrow K^{\R}+\delta K^{\R}$, and its eigenvalue, $k_\R(\alpha)\rightarrow k_\R(\alpha)+\delta k_\R(\alpha)$. The corresponding first-order shift $\delta\varkappa$ of the Lyapunov exponent can be found as follows:
\begin{equation}
k_\R(-\varkappa-\delta\varkappa)+\delta k_\R(-\varkappa)=1 \quad \Rightarrow \quad \delta\varkappa =\frac{\delta k_\R(-\varkappa)}{t_B}\,.
\end{equation}

Now, we will compute the branching time by a generating function method. Within the thin rung  approximation, the OTOC is given by a sum of diagrams with $n$ rungs (denoted by $F_n$),
\begin{equation}
\OTOC(t_1,t_2,t_3,t_4) =\sum_{n=0}^{\infty} F_n(t_1,t_2,t_3,t_4) \,.
\end{equation}
We may interpret $F_n/\OTOC$ as the ``probability''\footnote{In general, $F_n(t_1,t_2,t_3,t_4)/\OTOC(t_1,t_2,t_3,t_4)$ may not be positive.} for a $n$-rung ladder to appear in the $\OTOC$, and define the average number of rungs as 
\begin{equation}
\overline{n}(t_1,t_2,t_3,t_4)=\sum_{n=0}^{\infty}\frac{n F_n(t_1,t_2,t_3,t_4)}{\OTOC(t_1,t_2,t_3,t_4)} \,.
\end{equation}
To proceed, let us introduce a generating function with a parameter $\mu$ representing the chemical potential for rungs,
\begin{equation}
Z(\mu,t_1,t_2,t_3,t_4)=\sum_{n=0}^{\infty}e^{\mu n}F_n(t_1,t_2,t_3,t_4)=\sum_{n=0}^{\infty}
\begin{tikzpicture}[baseline={([yshift=0pt]current bounding box.center)}]
\draw[thick, decorate, decoration={snake, segment length=11pt, amplitude=2pt
}] (0pt,-15pt) -- (0pt,15pt);
\draw[thick, decorate, decoration={snake, segment length=11pt, amplitude=2pt
}] (18pt,-15pt) -- (18pt,15pt);
\draw[thick, decorate, decoration={snake, segment length=11pt, amplitude=2pt
}] (50pt,-15pt) -- (50pt,15pt);

\draw[thick] (-15pt,15pt)--(65pt,15pt);
\draw[thick] (-15pt,-15pt)--(65pt,-15pt);

\filldraw  (-15pt,15pt) circle (1pt) node[left] {\scriptsize $t_1$};
\filldraw  (-15pt,-15pt) circle (1pt) node[left] {\scriptsize$t_2$};
\filldraw  (65pt,15pt) circle (1pt) node[right] {\scriptsize$t_3$};
\filldraw  (65pt,-15pt) circle (1pt) node[right] {\scriptsize$t_4$};

\node at (30pt,-2pt) {\scriptsize $\ldots$};
\node at (-7.5pt,0pt) {\scriptsize $e^\mu$};
\node at (10.5pt,0pt) {\scriptsize $e^\mu$};
\node at (42.5pt,0pt) {\scriptsize $e^\mu$};
\node at (0pt,-20pt) {\scriptsize $1$};
\node at (18pt,-20pt) {\scriptsize $2$};
\node at (50pt,-20pt) {\scriptsize $n$};
\end{tikzpicture}\,;
\end{equation}
then the average number of rungs is the logarithmic derivative of the generating function, namely,
\begin{equation}
\overline{n}(t_1,t_2,t_3,t_4)=\partial_\mu \ln  Z(\mu,t_1,t_2,t_3,t_4)\big |_{\mu=0} \,.
\label{eqn: ave n}
\end{equation}
A useful observation is that 
the generating function $Z(\mu,t_1,t_2,t_3,t_4)$ can be determined by the same kinetic equation approach, with a weighted kernel $e^{\mu} K^{\R}$:
\begin{equation}
\int dt_5\,dt_6\, e^\mu K^{\R}(t_1,t_2,t_5,t_6)Z(\mu,t_5,t_6,t_3,t_4)
\approx Z(\mu,t_1,t_2,t_3,t_4)\,.
\end{equation}  
This suggests $Z(\mu,t_1,t_2,t_3,t_4)$ should have a similar form to the OTOC,
\begin{equation}
Z(\mu,t_1,t_2,t_3,t_4)=\frac{e^{\varkappa(\mu) (t_1+t_2-t_3-t_4)/2   }}{{C(\mu)}} \VF^\R(t_{12},\mu) \VF^\A (t_{34},\mu) \,,
\end{equation}
where $C(\mu)$ and $\VF^{\R/\A}(t,\mu)$ are smooth functions near $\mu=0$. Plugging the last equation into \eqref{eqn: ave n}, we get 
\begin{equation}\label{average n1}
\overline{n}(t_1,t_2,t_3,t_4)
= \varkappa'(0)\frac{t_1+t_2-t_3-t_4}{2}+(\text{non-growing part}). 
\end{equation}
The number $\varkappa(\mu)$ is determined by solving the eigenvalue equation $e^\mu k_\R(-\varkappa(\mu))=1$. Differentiating it with respect to $\mu$, we find
\begin{equation}\label{dkappa}
\varkappa'(0)= \frac{1}{k_\R'(-\varkappa)}=\frac{1}{t_B}\,.
\end{equation}
Practically speaking, this relation provides an alternative route to the branching time. The branching time is the change in the Lyapunov exponent when a weight $e^{\mu}$ is added to the retarded kernel $K^\R$. This method will be applied in following sections to prove a bound for the branching time and to estimate the branching time of weakly interacting systems. 

We would also like to comment on the meaning of equation \eqref{average n1}. Together with \eqref{dkappa}, it gives
\begin{equation}\label{average n}
\overline{n}(t_1,t_2,t_3,t_4)\approx \frac{t_1+t_2-t_3-t_4}{2t_B} \,,
\end{equation}
which is consistent with the interpretation of $t_B$ as the average rung separation. Note that \eqref{average n} is correct only for large time differences such that
\begin{equation}
\bar{n^2}-\bar{n}^2=\partial^2_\mu \ln Z(\mu,t_1,t_2,t_3,t_4)|_{\mu=0}\approx\varkappa''(0)\frac{t_1+t_2-t_3-t_4}{2} \ll \bar{n}^2\,,
\end{equation}
and therefore, the distribution of $n$ approaches a delta function.

\section{A bound on the branching time}
\label{sec: bound}

In this section, we present the main result of this paper: we show that within the  thin rung approximation, the following bound on the branching time holds for fermionic models:
\begin{equation}
\wideboxed{
t_B \varkappa \leq 2 \,.
}
\label{eqn: branching bound}
\end{equation}
We will prove it using the generating function trick with parameter $\mu$, and express the branching time as the $\mu$-derivative of $\varkappa$ via \eqref{dkappa}. Our proof involves two steps:
\begin{enumerate}
\item Relate $t_B$ to 
$\partial_{\omega}\phi$, where 
\begin{equation}
\phi\left(\omega+i\frac{\varkappa}{2}\right)= -\frac{i}{2} \ln \frac{\tG^{\R}(\omega+i \frac{\varkappa}{2})}{
\tG^{\A}(\omega- i \frac{\varkappa}{2})
}
\end{equation}
is the phase of the retarded Green function on the upper half plane.  
This step is completed by a 
Hellmann-Feynman type argument, which will be shown momentarily.
\item Derive a bound on $\partial_{\omega}\phi$. 

This second part is common to fermionic Green functions (i.e.\ the result does not rely on the kinetic equation or the thin rung approximation). For this reason, we have put the proof of the bound in a separate section as Appendix~\ref{appendix: proof}.
\end{enumerate}

We start with the deformed kernel $e^{\mu} K^R$ and the following equation (i.e.\ introduce a parameter $\mu$ to Eq.~\eqref{eqn: key equation})
 \begin{equation}\label{eqn: mu kinetic}
 e^\mu W\Big(\omega +  i \frac{\varkappa(\mu)}{2}\Big) \int  \tilde R(\omega-\omega') \tVF^{\R}(\omega',\mu) \frac{d\omega'}{2\pi} =\tVF^{\R} (\omega,\mu) \,.
 \end{equation}
 Since $W>0$ away from the real axis, we can equivalently write
\begin{equation}
e^\mu\int \tilde R(\omega-\omega')\tVF^\R(\omega',\mu)\frac{d\omega'}{2\pi} =
\frac{\tVF^{\R}(\omega,\mu)}{W\big(\omega+i\frac{\varkappa(\mu)}{2}\big)}\,.
\label{eqF}
\end{equation}
Using the property $\tilde R(\omega)=\tilde R(-\omega)^*$ (i.e.\ the reality of the rung function in the time representation, $R(t)=R(t)^*$ --- see Appendix~\ref{appendix: Rung} for a proof), we get the following formula for the complex conjugate of \eqref{eqF}:
\begin{equation}
e^\mu\int \tilde R(-\omega+\omega')\tVF^\R(\omega',\mu)^* \frac{d\omega'}{2\pi} =
\frac{\tVF^{\R}(\omega,\mu)^*}{W\big(\omega+i\frac{\varkappa(\mu)}{2}\big) }\,.
\label{eqFstar}
\end{equation}
We have also used the fact that $W$ is real. A neater way to write these formulas is to use bra $\langle \tVF^\R |$ and ket $|\tVF^\R\rangle$ with the Hermitian inner product given by the integral over $\frac{d\omega}{2\pi}$. Then $\tilde R$ may be regarded as a Hermitian operator with the matrix elements $\langle\omega|\tilde R|\omega'\rangle =\tilde R(\omega-\omega') =\tilde R(\omega'-\omega)^* =\langle\omega'|\tilde R|\omega\rangle^*$, while $W$ is a diagonal Hermitian operator. Thus, \eqref{eqF} and \eqref{eqFstar} become
\begin{equation}
e^\mu \tilde R |\tVF^\R\rangle = W^{-1} |\tVF^\R\rangle \,, \qquad 
e^\mu \langle \tVF^\R |\tilde R  =\langle \tVF^\R | W^{-1}  \,.
\label{eqn: vectors}
\end{equation}
Consequently, we have $
e^\mu \langle \tVF^\R |\tilde R |\tVF^\R\rangle = \langle\tVF^\R | W^{-1} |\tVF^\R\rangle$. 
Next, we run a Hellmann-Feynman type of argument: when we take the $\mu$ derivative of both sides of the equation $e^\mu \langle \tVF^\R |\tilde R |\tVF^\R\rangle = \langle\tVF^\R | W^{-1} |\tVF^\R\rangle$, the terms involving derivatives of $\langle\tVF^\R | $ and $|\tVF^\R\rangle $ cancel due to \eqref{eqn: vectors}. Therefore, only the derivatives of $e^\mu \tilde R$ and $W^{-1}$ survive, namely,
\begin{equation}
\Bigl\langle \tVF^\R \Big| \frac{d W^{-1}}{d \mu}  \Big|\tVF^\R\Bigr\rangle = 
e^\mu \bigl\langle \tVF^\R \big|\tilde R \big|\tVF^\R\bigr\rangle
= \bigl\langle \tVF^\R \big|W^{-1} \big|\tVF^\R\bigr\rangle   \,.
\end{equation} 
In the last step, we used equation \eqref{eqn: vectors} again. 
Recall that $W\bigl( \omega+i\frac{\varkappa}{2} \bigr) = \tG^{\R}\bigl(\omega +  i \frac{\varkappa}{2}\bigr) \tG^{\A}\bigl(\omega - i \frac{\varkappa}{2}\bigl) = 
\left|  \tG^{\R}\bigl(\omega +  i \frac{\varkappa}{2}\bigr)  \right|^2$ is the square of the magnitude of $\tG^\R$, whose derivative along the imaginary axis is related to the derivative of the phase of $\tG^\R$ along the real axis. More explicitly, we have the following formula:
\begin{equation}
\frac{d W^{-1}}{d \mu}  = W^{-1}\varkappa'(\mu)\, \partial_{\omega}\phi \Big(\omega+i\frac{\kap}{2} \Big) \,, \quad \text{where} \quad \phi \left(\omega+i\frac{\kap}{2} \right) :=  - \frac{i}{2}  \ln \frac{\tG^\R(\omega+i\frac{\varkappa}{2})}{\tG^\A(\omega-i\frac{\varkappa}{2})}  \,.
\end{equation}
Setting $\mu=0$, we obtain $t_B=\varkappa'(0)^{-1}$:
\begin{equation}
t_B= \frac{\Big\langle \tVF^\R \Big| W^{-1} \frac{d\phi }{d\omega}  \Big|\tVF^\R\Big\rangle }{\Big\langle \tVF^\R \Big| W^{-1}  \Big|\tVF^\R\Big\rangle } \quad \text{at} \quad \mu=0 \,.
\label{tbFormula}
\end{equation}
We would like to interpret the above expression for $t_B$ as the average value of the winding speed of $\phi$ with some probability distribution $P$. Such an interpretation is indeed possible:
\begin{equation}
t_B= \int d\omega P(\omega)  \frac{d\phi }{d\omega}\,,  \qquad
P(\omega) = \frac{1}{\calN} \left|\frac{\tVF^\R(\omega)}{\tG^\R(\omega+i\frac{\varkappa}{2})}\right|^2\,,
\label{eqn: tb prob}
\end{equation}
where $\calN=\int d\omega \left|\frac{\tVF^\R(\omega)}{\tG^\R(\omega+i\frac{\varkappa}{2})}\right|^2 $ is a normalization factor. 
Next, we will bound $t_B$ using a lemma that bounds the winding speed of $\phi$. 
\begin{defn}[Lemma.] The retarded/advanced Green function $G^{\R/\A}$ of an arbitrary Fermi system satisfies the following inequality for all $\omega\in \RR$ and $s\in\RR^+$:
\begin{equation}
-\frac{i}{2} \partial_\omega \ln \frac{\tG^\R(\omega+i s)}{\tG^\A(\omega-is) }\leq\frac{1}{s}\,.
\label{ieq}
\end{equation}
The proof is given in the appendix~\ref{appendix: proof}. 
\end{defn}
Applying the Lemma with $s=\frac{\varkappa}{2}$ to \eqref{eqn: tb prob}, we conclude that $t_B \leq \frac{2}{\varkappa}$, which is equivalent to the bound \eqref{eqn: branching bound}.\medskip

Now we make a few comments on the bound. 
\begin{enumerate}
\item It is not clear if one can get arbitrarily close to saturating the bound $t_B \kap\leq 2$. The best example we know is the SYK model at strong coupling (see Eq.~(52) of Ref.~\cite{gu2019relation}), where
\begin{equation}
t_B \varkappa = \pi \cot (2\pi \Delta) - \frac{1}{2\Delta} - \frac{1}{2\Delta-1} - \frac{1}{2\Delta-2} \in \left( 0, 3/2 \right) \,.
\end{equation}

The equal sign in the Lemma is achievable if the spectral function is a delta function; for example, if $A(\omega)=2\pi \delta(\omega)$, then the equality is attained at $\omega=0$. Note that in order to get an equal sign in the branching time bound, $t_B\varkappa=2$, one needs the bound in the Lemma to be tight whenever the ratio $\tVF^\R(\omega)/\tG^\R(\omega+i\frac{\varkappa}{2})$ is non-zero. On the other hand, the equality in the Lemma can not hold for all $\omega$ because the total change of the phase difference between $\tG^\R$ and $\tG^\A$ is fixed, namely,
\begin{equation}
\int_{-\infty}^{+\infty}  \left(
-\frac{i}{2} \partial_\omega \ln \frac{\tG^\R(\omega+i s)}{\tG^\A(\omega-is) }  
\right) d\omega=\pi\,.
\end{equation}
Therefore, we expect that the equality in \eqref{ieq} can be achieved only at isolated points. On the other hand, $\VF^\R(t)$ tends to zero at $t\to\pm\infty$, and therefore, its Fourier transform can not be a sum of delta functions. Combining these two observations, we expect that the branching time bound can not be saturated in any physical model.

For a more rigorous argument along similar lines, let us assume that $G^{\R}(t,0)$ decays as $e^{-\Gamma t/2}$ as $t$ goes to infinity. In the frequency representation, this means that $\tG^{\R}(z)$ is analytic for $\Im z>-\frac{\Gamma}{2}$. As a result, the lemma proved in appendix~\ref{appendix: proof} can be strengthened as follows:
\begin{equation}
-\frac{ i}{2}\partial_\omega\ln \frac{\tG^\R(\omega+i\frac{\varkappa}{2})}{\tG^\A(\omega-i\frac{\varkappa}{2})} \leq\frac{2}{\varkappa+\Gamma} \,,
\end{equation}
which consequently provides a tighter upper bound on the branching time,
\begin{equation}\label{weak coupling bound}
t_B \leq \frac{2}{\varkappa+\Gamma}\,.
\end{equation}

\item The key assumption for the bound to hold is the thin rung approximation for OTOCs. This condition might be violated, for example, in matrix models. 

\item Our proof works only for fermionic systems, since the proof of the Lemma relies on the positivity of the spectral function, which is only true for fermionic Green functions. 

\item The proof can be generalized to the higher-dimensional case under mild assumptions. The only change is to add momentum arguments to the vertex and Green functions and to integrate over the momentum whenever there is a loop in the diagram. The conclusion is that the bound $t_B \kap\leq 2$ still holds, provided  the center-of-mass momentum of the exponentially growing mode is zero. One also needs the identity $\tilde R(\vec{p}-\vec{q},\omega-\omega') =\tilde R(\vec{q}-\vec{p},\omega'-\omega)^*$, which is analogous to the multi-flavor case discussed below.

\item The bound and its proof also generalize to multi-flavor Fermi system, where we need to introduce flavor indices for the rung function $\tilde R_{ab}$, the Green functions $\tG^{\R,\A}_{a}$, and vertex functions $\tVF^{\R,\A}_{a}$. Here, we have assumed that the Green function is diagonal in the flavor basis, i.e.\ $\tG^{\R,\A}_{ab}=\delta_{ab} \tG^{\R,\A}_a$. Diagrammatically, the rung function looks like this:
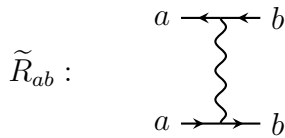
\begin{equation*}
\tilde R_{ab}: \qquad
\begin{tikzpicture}[baseline={([yshift=-4pt]current bounding box.center)}]
\draw[mid arrow, thick] (15pt,20pt)--(0pt,20pt);
\draw[mid arrow, thick] (0pt,20pt)--(-15pt,20pt);
\draw[mid arrow, thick] (-15pt,-20pt)--(0pt,-20pt);
\draw[mid arrow, thick] (0pt,-20pt)--(15pt,-20pt);
\draw[thick, decorate, decoration={snake, segment length=11pt, amplitude=2pt
}] (0pt,20pt) -- (0pt,-20pt);
\node[left] at (-15pt,20pt) {$a$};
\node[right] at (15pt,20pt) {$b$};
\node[left] at (-15pt,-20pt) {$a$};
\node[right] at (15pt,-20pt) {$b$};
\end{tikzpicture} 
\end{equation*}
Note that we require the incoming and outgoing flavors on each side to be the same. For SYK-like models with disorder, this assumption is true if the disorder is flavor-diagonal, e.g.\ if $\overline{J_{abcd}J_{a'b'c'd'}}\propto \delta_{aa'}\delta_{bb'}\delta_{cc'}\delta_{dd'}$. A natural generalization of the reality condition for $R$ is Hermiticity, namely, $\tilde R_{ab}(\omega-\omega')=\tilde R_{ba}(-\omega+\omega')^*$, see appendix~\ref{appendix: Rung} for a proof. The derivation of the inequality $t_B \kap\leq 2$ requires minor modifications but remains correct.

\end{enumerate}

\section{Branching time at weak coupling}
\label{sec: weak}

In this section, we investigate the special case where the Green function has a quasiparticle pole, which is expected at weak coupling. 
To be concrete, let us take the Majorana SYK model as an example and comment on the general case and higher dimensions later. In the limit of weak interaction, i.e.\ at temperatures $\beta^{-1}\gg J$, we may approximate the retarded Green function by that of a quasiparticle with zero energy and lifetime $\tau_{\text{qp}}=1/\Gamma$, where $\Gamma\sim J$:
\begin{equation}\label{GRquasi}
\tG^\R(\omega) \approx \frac{1}{\omega+i\Gamma/2} \,.
\end{equation}
In general, $\tG^\R$ could have multiple poles characterized by decay rates $\Gamma$ of the same order of magnitude. As a consequence, the calculation of $t_B$ based on \eqref{GRquasi} in the following subsections will only give an order-of-magnitude estimate. See appendix~\ref{appendix: gamma} for more discussions on the estimation of $\Gamma$.

\subsection{A bound state problem}


We have previously mentioned the bound $t_B \leq \frac{2}{\varkappa+\Gamma}$. This result, along with an actual estimate for $t_B$, can also be derived using the intuition from solving a bound state problem in ordinary quantum mechanics, similar to the large-$q$ SYK discussed in Ref.~\cite{MS16-remarks}. We would like to explain the quantum mechanical interpretation here, which will also be useful for later discussions of various approximation methods. 

Let us start with the kinetic equation \eqref{eqF}  with parameter $\mu$  and insert the quasiparticle ansatz \eqref{GRquasi} into $W=|\tG^\R|^2$, which yields $W=\frac{1}{\omega^2+(\frac{\varkappa+\Gamma}{2})^2}$ so that equation becomes
\begin{equation}
e^\mu\int \tilde R(\omega-\omega')\tVF^\R(\omega',\mu)\frac{d\omega'}{2\pi} = \left( \omega^2 + \frac{(\varkappa+\Gamma)^2}{4} \right) \tVF^{\R}(\omega,\mu)\,.
\end{equation}
In the time domain, the above equation turns into a second order differential equation,
\begin{equation}\label{eqn: bound state}
\left( - \partial_t^2-  e^{\mu} R(t)\right) \VF^{\R}(t,\mu)  =-  \frac{(\varkappa+\Gamma)^2}{4} \VF^{\R}(t,\mu)  \,,
\end{equation}
which can be further interpreted as a Schrodinger equation with potential $V=-  e^{\mu} R(t)$ and total energy $E=-  \frac{(\varkappa+\Gamma)^2}{4}<0$ (i.e.\ a bound state problem). Using the Hellmann-Feynman theorem, we get the potential energy at $\mu=0$ by taking the $\mu$-derivative of the total energy, i.e.
\begin{equation}
E_V= - \frac{\varkappa+\Gamma}{2} \varkappa'(0) = - \frac{\varkappa+\Gamma}{2t_B} \,. 
\end{equation}
Note that the kinetic energy $E_K$ is always non-negative, and hence, $E=E_K+E_V \geq E_V $. Expressing $E$ and $E_V$ in terms of $\varkappa$, $\Gamma$, and $t_B$, we get
\begin{equation}
\frac{(\varkappa+\Gamma)^2}{4} \leq \frac{\varkappa+\Gamma}{2t_B}   \quad \text{i.e.} ~~ t_B \leq \frac{2}{\varkappa+\Gamma}\,,
\end{equation}
which reproduces the bound at weak coupling \eqref{weak coupling bound}.

\subsection{Comments on various approximation methods}
\label{section: comments}

The bound state interpretation provides a useful clue for various  approximation methods. We will start with the zero range potential approximation (see Ref.~\cite{demkov2013zero} for an exposition) for the rung function (potential term in the QM interpretation) and derive an approximate relation between the branching time $t_B$, the Lyapunov exponent $\varkappa$, and the quasiparticle decay rate $\Gamma$:
\begin{equation}
\wideboxed{
\varkappa \approx \frac{1}{t_B} - \Gamma \,.
}
\label{eqn: branching relation}
\end{equation}
In other words, the branching time is half of its upper bound \eqref{weak coupling bound}. We also comment on a widely used approximation in the literature, the delta function approximation in the kinetic equation introduced by Stanford \cite{Stanford:2015owe}, and show that it leads to the same relation.

\begin{defn}[Zero-range potential approximation for the bound state energy.] 
For the bound state problem \eqref{eqn: bound state}, the simplest approximation is to replace the potential with a delta-function: 
\begin{equation}
R(t) \,\rightarrow\, \tilde R(0) \delta	(t) \quad \text{with} \quad \tilde R(0) = \int_{-\infty}^{+\infty} R(t) dt \,.
\end{equation}
Then the wave function and the corresponding bound state energy 
$E= - \frac{(\varkappa+\Gamma)^2}{4} $
are expressed as follows
\begin{equation}
\VF^\R(t,\mu) \sim  e^{- \lambda |t|} \,, \quad \lambda = \frac{e^\mu \tilde R(0)}{2} \,, \quad E = -\lambda^2 \,.
\label{eqn: solution}
\end{equation}
Therefore, we have $\varkappa(\mu) = e^\mu \tilde R(0) - \Gamma$ and 
\begin{equation}
\varkappa = \tilde R(0)-\Gamma, \quad t_B = \varkappa'(0)^{-1} = \tilde R(0)^{-1}\,.
\end{equation}
Eliminating $\tilde R(0)$, we get the relation \eqref{eqn: branching relation}:
\begin{equation}
\varkappa = \frac{1}{t_B} - \Gamma \,.
\end{equation}
This formula shows that at weak coupling, branching is essential for scrambling to occur, which is in contrast to the strong coupling limit, where the branching slows down the scrambling. For example, in the SYK model with $J\gg 1$, the deviation of the Lyapunov exponent from its maximal value is
\begin{equation}
\delta \varkappa = 1- \varkappa \sim \frac{1}{J t_B} \qquad \text{(in SYK at strong coupling)} \,.
\end{equation}
The contrast suggests that there are actually two different mechanisms for scrambling, one at weak coupling and another at strong coupling (corresponding, respectively, to high and low temperature) \cite{kitaev2017near,kitaev2018kitp}. 

As to the validity of the zero-range potential approximation we have used, it should work when the range of the potential $R(t)$ (denoted as $t_0$) is much smaller than the size of the wave packet given by $\VF^\R(t,\mu)$ in \eqref{eqn: solution}, namely
\begin{equation}
t_0 \ll  \tilde R(0)^{-1} = t_B \,.
\label{eqn: criterior}
\end{equation}
For example, in appendix~\ref{appendix: brownian}, we show that  this criterion is satisfied for the Brownian SYK\cite{Saad:2018bqo,Sunderhauf:2019djv}, where the rung function is indeed a delta function, and therefore, the relation \eqref{eqn: branching relation} is exact. On the other hand, there is also a weakly coupled model where this approximation (and the approximations below) are not accurate. 
In appendix~\ref{appendix: large q}, we show that for the (regular) large $q$ SYK at weak coupling, there is an order $1$ prefactor discrepancy between the approximated and exact results. 

\end{defn}

\begin{defn}[Delta function approximation in the kinetic equation.] 
The zero range potential approximation for the bound state problem above is similar to the type of approximation used in \cite{Stanford:2015owe} by Stanford, where the product of retarded and advanced Green functions (which is denoted as $W$ in this paper) is approximated by a delta function:
\begin{equation}
W^{\R}\Big(\omega +  i \frac{\varkappa(\mu)}{2}\Big) 
\approx
\frac{1}{\omega^2 + \bigl(\frac{\varkappa+\Gamma}{2}\bigl)^2}= \underbrace{\frac{\varkappa+\Gamma}{\omega^2+\bigl(\frac{\varkappa+\Gamma}{2}\bigr)^2} }_{\approx 2\pi \delta(\omega)}
\cdot
 \frac{1}{\varkappa+\Gamma} 
 \,.
\label{eqn: approx}
\end{equation}
In the last step, the Lorentzian function is replaced by the delta function.\footnote{For a slightly different explanation, see Ref.~\cite{Stanford:2015owe}}  Therefore, we have a simplified equation,
 \begin{equation}
 e^\mu 2\pi \delta(\omega)  \int  \tilde R(\omega-\omega') \tVF^{\R}(\omega',\mu) \frac{d\omega'}{2\pi} \approx (\varkappa+\Gamma) \tVF^{\R} (\omega,\mu) \,.
 \end{equation}
The appearance of the delta function supplies additional convenience: the vertex function $\tVF^\R(\omega, \mu)$ should also be proportional to $\delta (\omega )$ in this approximation. Thus, we have the following relation,
\begin{equation}
\varkappa(\mu)+\Gamma = e^\mu \tilde R(0) \,,
\end{equation}
which is identical to the result obtained using the zero range potential approximation, and therefore, entails the same conclusion,
\begin{equation}
\varkappa= \frac{1}{t_B} -\Gamma\,.
\end{equation}
We would like to make a few additional comments about this approach:
\begin{enumerate}
\item The approximation in \eqref{eqn: approx} is valid when $\varkappa+\Gamma \ll \omega_0$, where $\omega_0$ is the frequency scale below which the rung function $\tilde R(\omega)$ is almost constant. This is merely the frequency space version of the criterion \eqref{eqn: criterior}.

\item The computations here  generalize to higher dimensions and the multi-flavor case straightforwardly, where the retarded Green function has the following form 
\begin{equation}
\tG^\R_a(\vec{p}, \omega)= \frac{1}{\omega-\varepsilon_a(\vec{p})+i \Gamma_a(\vec{p})/2} \,.
\end{equation}
The subscript $a$ labels the quasiparticle flavor, and $\vec{p}$ is the momentum vector. In this case, we need to consider a generalized version of equation \eqref{eqn: mu kinetic}, namely,
 \begin{equation}
 e^\mu W_a^{\R}\Big(\vec{p},  \omega +  i \frac{\varkappa}{2}\Big)   
\sum_b 
 \int  \tilde R_{ab}(\vec{p}-\vec{q}, \omega-\omega') \tVF_b^{\R}(\vec{q},\omega',\mu) \frac{d\omega'}{2\pi} \frac{d^d \vec{q}}{(2\pi)^d} =\tVF_a^{\R} (\vec{p},\omega,\mu) \,.
 \end{equation}
Next, we use the following approximation in the above equation,
\begin{equation} 
W_a^{\R}\Big(\vec{p},  \omega +  i \frac{\varkappa}{2}\Big) \approx \frac{2\pi \delta(\omega-\varepsilon_a(\vec{p}))}{\varkappa+\Gamma_a(\vec{p})}\,,
\end{equation}
which leads to the following result:
\begin{equation}
\varkappa(\mu) =\langle  e^{\mu}{\tilde R} - \Gamma \rangle\,.
\end{equation}
Here ${\tilde R}$ and ${\Gamma}$ are understood as matrices in the flavor ($a,b$) and momentum ($\vec{p},\vec{q}$) indices, namely,
\begin{equation}
{\tilde R}_{a,\vec{p};b,\vec{q}} = \tilde R_{ab}(\vec{p}-\vec{q},\eps(\vec{p})-\eps(\vec{q})) \,, \quad \Gamma_{a,\vec{p};b,\vec{q}}= \delta_{ab} \delta(\vec{p}-\vec{q})\, \Gamma_a(\vec{p}) 
\end{equation}
and the expectation value $\langle \cdot \rangle$ is taken on the eigenvector with largest eigenvalue $\varkappa(\mu)$. Then the branching time is given by the Hellmann-Feynman theorem, with $1/t_B=\langle \tilde {R} \rangle$ evaluated on the same eigenvector, which leads to the following relation:
\begin{equation}
\varkappa= \frac{1}{t_B} - \langle {\Gamma} \rangle\,.
\end{equation}
\end{enumerate}
\end{defn}

\section{Summary and discussion}

Our main result is the bound on the branching time,  
$
\varkappa t_B \leq 2
$, 
for a large class of SYK-like models, which may be interpreted as a warning sign against naive attempts to obtain a holographic model with bulk locality. On the other hand, the assumptions used to prove the bound might serve as a guide for the search of desired models in a broader class. A general retarded kernel contains four matrix indices and a rung function of four time variables:
\begin{equation}
K^\R_{ab;cd}= \quad
\begin{tikzpicture}[baseline={([yshift=-4pt]current bounding box.center)},
  vertex1/.style={rectangle,draw=black,thick,fill=lightgray,minimum size=18pt,align=center}]
  \node[vertex1] (R) at (0pt,0pt) { \\  \\  };
  \node at (0pt,0pt) {$R$};
\draw[mid arrow, thick] (20pt,18.5pt)--(9pt,18.5pt);
\draw[thick] (-9pt,18.5pt)--(9pt,18.5pt);
\draw[mid arrow, thick] (-9pt,18.5pt)--(-40pt,18.5pt);

\draw[mid arrow, thick] (-40pt,-18.5pt)--(-9pt,-18.5pt);
\draw[thick] (-9pt,-18.5pt)--(12pt,-18.5pt);
\draw[mid arrow, thick] (12pt,-18.5pt)--(20pt,-18.5pt);
\node[left] at (-40pt,18.5pt) {$a$};
\node[right] at (20pt,18.5pt) {$c$};
\node[left] at (-40pt,-18.5pt) {$b$};
\node[right] at (20pt,-18.5pt) {$d$};
\end{tikzpicture}\quad, \qquad
R_{ab;cd}(t_1,t_2,t_3,t_4)= \quad
\begin{tikzpicture}[baseline={([yshift=-4pt]current bounding box.center)},
  vertex1/.style={rectangle,draw=black,thick,fill=lightgray,minimum size=18pt,align=center}]
  \node[vertex1] (R) at (0pt,0pt) { \\  \\  };
  \node at (0pt,0pt) {$R$};
\draw[thick] (15pt,18.5pt)--(-15pt,18.5pt);
\draw[ thick] (-15pt,-18.5pt)--(15pt,-18.5pt);
\filldraw (-15pt,18.5pt) circle (1pt) node[left] {$a,t_1$};
\filldraw  (15pt,18.5pt) circle (1pt) node[right] {$c,t_3$};
\filldraw (-15pt,-18.5pt) circle (1pt) node[left]{$b,t_2$};
\filldraw  (15pt,-18.5pt) circle (1pt) node[right]{$d,t_4$};
\end{tikzpicture}\quad.
\end{equation}
Here all indices are different. Each operator can be bosonic or fermionic, as long as the total fermion parity is even. The rung function $R_{ab;cd}(t_1,t_2,t_3,t_4)$ is the sum of general 2-particle irreducible (2PI) diagrams. To derive the bound, we made the following assumptions:

\begin{enumerate}
\item  The conservation of flavor singlet. We assumed that we could restrict to the subspace where the Green functions on the two parallel rails for the retarded kernel are fermionic and with the same flavor index. In other words, we require the ``scramblon'' that mediates the propagation of chaos to be a singlet in the fermion flavor: 
\begin{equation}
K^\R_{ab}=K^\R_{aa;bb}: \qquad
\begin{tikzpicture}[baseline={([yshift=-4pt]current bounding box.center)},
  vertex1/.style={rectangle,draw=black,thick,fill=lightgray,minimum size=18pt,align=center}]
  \node[vertex1] (R) at (0pt,0pt) { \\  \\  };
  \node at (0pt,0pt) {$R$};
\draw[mid arrow, thick] (20pt,18.5pt)--(9pt,18.5pt);
\draw[thick] (-9pt,18.5pt)--(9pt,18.5pt);
\draw[mid arrow, thick] (-9pt,18.5pt)--(-40pt,18.5pt);

\draw[mid arrow, thick] (-40pt,-18.5pt)--(-9pt,-18.5pt);
\draw[thick] (-9pt,-18.5pt)--(12pt,-18.5pt);
\draw[mid arrow, thick] (12pt,-18.5pt)--(20pt,-18.5pt);
\node[left] at (-40pt,18.5pt) {$a$};
\node[right] at (20pt,18.5pt) {$b$};
\node[left] at (-40pt,-18.5pt) {$a$};
\node[right] at (20pt,-18.5pt) {$b$};
\end{tikzpicture}
\end{equation}
In SYK-like models, this conservation owes to the disorder averaging and the condition that the disorder is flavor-diagonal. 

\item The model is purely fermionic. We used the fact that the spectral function is positive semi-definite, which is not true for bosonic operators. It is then interesting to study models with both bosons and fermions \cite{MSW17}, and ask whether similar bounds exists. 

\item The thin rung approximation requires that the rung function only depend on two times (and further, due to the time translation symmetry, it only depends on the time difference of the two ends): 
\begin{equation}
R_{aa;bb}(t_1,t_2,t_3,t_4)=R_{ab}(t_1,t_2)\delta(t_{13})\delta(t_{24}): \quad
\begin{tikzpicture}[baseline={([yshift=-4pt]current bounding box.center)}]
\draw[thick] (15pt,18.5pt)--(-15pt,18.5pt);
\draw[ thick] (-15pt,-18.5pt)--(15pt,-18.5pt);
\draw[thick, decorate, decoration={snake, segment length=11pt, amplitude=2pt
}] (0pt,18.5pt) -- (0pt,-18.5pt);
\filldraw (-15pt,18.5pt) circle (1pt) node[left] {$a,t_1$};
\filldraw  (15pt,18.5pt) circle (1pt) node[right] {$b,t_3$};
\filldraw (-15pt,-18.5pt) circle (1pt) node[left]{$a,t_2$};
\filldraw  (15pt,-18.5pt) circle (1pt) node[right]{$b,t_4$};
\end{tikzpicture}
\end{equation}
It would be interesting to study models with more complicated rung functions. As an example, we could consider a matrix model, where the 2PI diagrams can be nontrivial:
\begin{equation}
R_{ab}(t_1,t_2;t_3,t_4)=\begin{tikzpicture}[baseline={([yshift=-4pt]current bounding box.center)}]
\draw[thick] (20pt,20pt)--(7pt,20pt);
\draw[thick] (7pt,-20pt)--(20pt,-20pt);
\draw[thick] (3pt,20pt)--(-10pt,20pt);
\draw[ thick] (-10pt,-20pt)--(3pt,-20pt);

\draw[thick] (7pt,-20pt)--(7pt,20pt);
\draw[thick] (3pt,-20pt)--(3pt,20pt);

\filldraw (-10pt,20pt) circle (1pt) node[left] {$a,t_1$};
\filldraw  (20pt,20pt) circle (1pt) node[right] {$b,t_3$};
\filldraw (-10pt,-20pt) circle (1pt) node[left]{$a,t_2$};
\filldraw  (20pt,-20pt) circle (1pt) node[right]{$b,t_4$};
\end{tikzpicture} +
\begin{tikzpicture}[baseline={([yshift=-4pt]current bounding box.center)}]
\draw[thick] (15pt,20pt)--(10pt,20pt);
\draw[thick] (6pt,20pt)--(-6pt,20pt);
\draw[thick] (-15pt,20pt)--(-10pt,20pt);

\draw[thick] (15pt,-20pt)--(2pt,-20pt);
\draw[thick] (-15pt,-20pt)--(-2pt,-20pt);

\draw[thick] (0pt,4pt)--(6pt,20pt);
\draw[thick] (0pt,4pt)--(-6pt,20pt);

\draw[thick] (2pt,0pt)--(10pt,20pt);
\draw[thick] (-2pt,0pt)--(-10pt,20pt);

\draw[thick] (-2pt,-20pt)--(-2pt,0pt);
\draw[thick] (2pt,-20pt)--(2pt,0pt);

\filldraw (-15pt,20pt) circle (1pt) node[left] {$a,t_1$};
\filldraw  (15pt,20pt) circle (1pt) node[right] {$b,t_3$};
\filldraw (-15pt,-20pt) circle (1pt) node[left]{$a,t_2$};
\filldraw  (15pt,-20pt) circle (1pt) node[right]{$b,t_4$};
\end{tikzpicture} 
+
\begin{tikzpicture}[baseline={([yshift=-4pt]current bounding box.center)}]
\draw[thick] (15pt,-20pt)--(10pt,-20pt);
\draw[thick] (6pt,-20pt)--(-6pt,-20pt);
\draw[thick] (-15pt,-20pt)--(-10pt,-20pt);

\draw[thick] (15pt,20pt)--(2pt,20pt);
\draw[thick] (-15pt,20pt)--(-2pt,20pt);

\draw[thick] (0pt,-4pt)--(6pt,-20pt);
\draw[thick] (0pt,-4pt)--(-6pt,-20pt);

\draw[thick] (2pt,0pt)--(10pt,-20pt);
\draw[thick] (-2pt,0pt)--(-10pt,-20pt);

\draw[thick] (-2pt,20pt)--(-2pt,0pt);
\draw[thick] (2pt,20pt)--(2pt,0pt);

\filldraw (-15pt,20pt) circle (1pt) node[left] {$a,t_1$};
\filldraw  (15pt,20pt) circle (1pt) node[right] {$b,t_3$};
\filldraw (-15pt,-20pt) circle (1pt) node[left]{$a,t_2$};
\filldraw  (15pt,-20pt) circle (1pt) node[right]{$b,t_4$};
\end{tikzpicture} 
+ \cdots
\end{equation}
Here, the double lines represent the matrix fields that mediate the interaction between fermions.

\end{enumerate}
It would also be interesting to understand logical connection between these conditions and bulk non-locality. 

Another motivation of our work was to investigate the relation between scrambling and branching. There are two distinct behaviors: for the SYK model at strong coupling, the branching (occurring at rate $1/t_B$) slows down the scrambling~\cite{gu2019relation},
 \begin{equation}
\varkappa \approx 1-\frac{\text{const}}{J t_B} \,,
\end{equation}
while in this paper, we have found the opposite tendency at weak coupling:
\begin{equation}
\kap \approx \frac{1}{t_B} - \Gamma \,.
\end{equation}
The contrast may be attributed to different modes of scrambling, coherent at low temperatures vs.\ incoherent at high temperatures~\cite{kitaev2017near,kitaev2018kitp}.

\section*{Acknowledgments}

We acknowledge Hui Zhai for collaboration in the early stages of this project. We thank Douglas Stanford for helpful discussions.
Y.G. is supported by
the Gordon and Betty Moore Foundation EPiQS Initiative through Grant (GBMF-4306), and 
by the Simons Foundation through the ``It from Qubit” program". 
A.K. 
is supported by the Simons Foundation under grant 376205 and through the ``It from Qubit'' 
program, as well as by the Institute of Quantum Information and Matter, a NSF Frontier 
center funded in part by the Gordon and Betty Moore Foundation. 
P.Z. 
acknowledges support from the Walter Burke Institute for Theoretical Physics at Caltech.

\appendix

\section{An alternative derivation of the ladder identity}
In Ref.~\cite{gu2019relation}, the following identity relating $t_B$, the Lyapunov exponent $\varkappa$, and the prefactor $C$, 
\begin{equation}
\label{eqn: ladder identity}
\wideboxed{
\frac{2 N \cos \frac{\varkappa \pi}{2}}{C} \cdot t_B \cdot \left( \VF^\A,\VF^\R \right) = 1 \,, \quad \text{where} \quad \left( \VF^\A,\VF^\R \right) := \int \VF^{\R} (t) R(t) \VF^{\A}(t) dt\,,
}
\end{equation}
was derived using a cut-and-glue consistency condition. Schematically, the procedure is to decompose a long ladder diagram into two shorter ladders glued by a ``box'' (formed by two rungs and two horizontal rails). The branching time $t_B$ appears as the typical size of the box in the aforementioned derivation. 

\begin{figure}[t]
\center
\begin{equation*}
\begin{tikzpicture}[baseline={([yshift=-4pt]current bounding box.center)}]
\draw[thick, decorate, decoration={snake, segment length=11pt, amplitude=2pt
}] (0pt,-15pt) -- (0pt,15pt);
\draw[blue,thick, decorate, decoration={snake, segment length=11pt, amplitude=2pt
}] (20pt,-15pt) -- (20pt,15pt);
\draw[thick, decorate, decoration={snake, segment length=11pt, amplitude=2pt
}] (40pt,-15pt) -- (40pt,15pt);
\draw[thick] (-10pt,15pt)-- (50pt,15pt);
\draw[thick] (-10pt,-15pt)-- (50pt,-15pt);
\end{tikzpicture}
\quad := \quad 
\begin{tikzpicture}[baseline={([yshift=-4pt]current bounding box.center)}]
\draw[thick, decorate, decoration={snake, segment length=11pt, amplitude=2pt
}] (0pt,-15pt) -- (0pt,15pt);
\draw[thick, decorate, decoration={snake, segment length=11pt, amplitude=2pt
}] (40pt,-15pt) -- (40pt,15pt);
\draw[thick] (-10pt,15pt)-- (50pt,15pt);
\draw[thick] (-10pt,17pt)-- (50pt,17pt);
\filldraw (-10pt,16pt) circle (1pt);
\draw[thick] (-10pt,-15pt)-- (50pt,-15pt);
\draw[thick] (-10pt,-17pt)-- (50pt,-17pt);
\filldraw (-10pt,-16pt) circle (1pt);
\draw[thick] (50pt,-17pt)-- (50pt,-30pt);
\draw[thick] (50pt,17pt)-- (50pt,30pt);
\draw[thick] (50pt,15pt)-- (50pt,-15pt);
\draw[blue,thick, decorate, decoration={snake, segment length=11pt, amplitude=2pt
}] (20pt,-15pt) -- (20pt,17pt);
\filldraw[blue] (20pt,-15pt) circle (1pt);
\filldraw[blue] (20pt,17pt) circle (1pt);
\end{tikzpicture}
\quad+\quad
\begin{tikzpicture}[baseline={([yshift=-4pt]current bounding box.center)}]
\draw[thick, decorate, decoration={snake, segment length=11pt, amplitude=2pt
}] (0pt,-15pt) -- (0pt,15pt);
\draw[thick, decorate, decoration={snake, segment length=11pt, amplitude=2pt
}] (40pt,-15pt) -- (40pt,15pt);
\draw[thick] (-10pt,15pt)-- (50pt,15pt);
\draw[thick] (-10pt,17pt)-- (50pt,17pt);
\filldraw (-10pt,16pt) circle (1pt);
\draw[thick] (-10pt,-15pt)-- (50pt,-15pt);
\draw[thick] (-10pt,-17pt)-- (50pt,-17pt);
\filldraw (-10pt,-16pt) circle (1pt);
\draw[thick] (50pt,-17pt)-- (50pt,-30pt);
\draw[thick] (50pt,17pt)-- (50pt,30pt);
\draw[thick] (50pt,15pt)-- (50pt,-15pt);
\draw[blue,thick, decorate, decoration={snake, segment length=11pt, amplitude=2pt
}] (20pt,-15pt) -- (20pt,15pt);
\filldraw[blue] (20pt,-15pt) circle (1pt);
\filldraw[blue] (20pt,15pt) circle (1pt);
\end{tikzpicture}
\quad+\quad
\begin{tikzpicture}[baseline={([yshift=-4pt]current bounding box.center)}]
\draw[thick, decorate, decoration={snake, segment length=11pt, amplitude=2pt
}] (0pt,-15pt) -- (0pt,15pt);
\draw[thick, decorate, decoration={snake, segment length=11pt, amplitude=2pt
}] (40pt,-15pt) -- (40pt,15pt);
\draw[thick] (-10pt,15pt)-- (50pt,15pt);
\draw[thick] (-10pt,17pt)-- (50pt,17pt);
\filldraw (-10pt,16pt) circle (1pt);
\draw[thick] (-10pt,-15pt)-- (50pt,-15pt);
\draw[thick] (-10pt,-17pt)-- (50pt,-17pt);
\filldraw (-10pt,-16pt) circle (1pt);
\draw[thick] (50pt,-17pt)-- (50pt,-30pt);
\draw[thick] (50pt,17pt)-- (50pt,30pt);
\draw[thick] (50pt,15pt)-- (50pt,-15pt);
\draw[blue,thick, decorate, decoration={snake, segment length=11pt, amplitude=2pt
}] (20pt,-17pt) -- (20pt,17pt);
\filldraw[blue] (20pt,-17pt) circle (1pt);
\filldraw[blue] (20pt,17pt) circle (1pt);
\end{tikzpicture}
\quad+\quad
\begin{tikzpicture}[baseline={([yshift=-4pt]current bounding box.center)}]
\draw[thick, decorate, decoration={snake, segment length=11pt, amplitude=2pt
}] (0pt,-15pt) -- (0pt,15pt);
\draw[thick, decorate, decoration={snake, segment length=11pt, amplitude=2pt
}] (40pt,-15pt) -- (40pt,15pt);
\draw[thick] (-10pt,15pt)-- (50pt,15pt);
\draw[thick] (-10pt,17pt)-- (50pt,17pt);
\filldraw (-10pt,16pt) circle (1pt);
\draw[thick] (-10pt,-15pt)-- (50pt,-15pt);
\draw[thick] (-10pt,-17pt)-- (50pt,-17pt);
\filldraw (-10pt,-16pt) circle (1pt);
\draw[thick] (50pt,-17pt)-- (50pt,-30pt);
\draw[thick] (50pt,17pt)-- (50pt,30pt);
\draw[thick] (50pt,15pt)-- (50pt,-15pt);
\draw[blue,thick, decorate, decoration={snake, segment length=11pt, amplitude=2pt
}] (20pt,-17pt) -- (20pt,15pt);
\filldraw[blue] (20pt,-17pt) circle (1pt);
\filldraw[blue] (20pt,15pt) circle (1pt);
\end{tikzpicture}
\end{equation*}
\caption{To explain why one of the OTOCs in \eqref{alt_derivation} is retarded, we draw the double Keldysh contour with time going left\protect\footnotemark\ and consider four ways of attaching the blue rung to it. The choice of attachment points affects the value of the left part of the diagram so that the four terms add up to the retarded OTOC. The right part is equal to the ordinary OTOC in all cases.} 
\label{fig: diag eqn}
\end{figure}
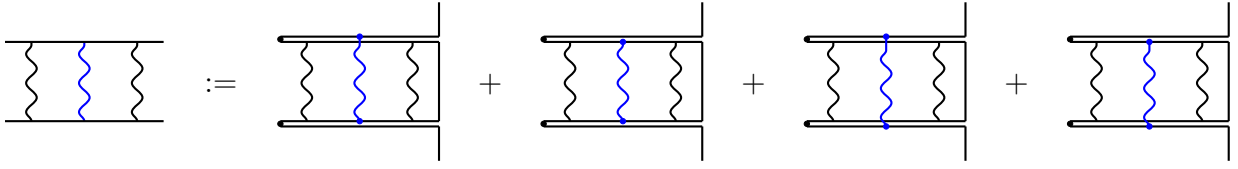

Now, we give an alternative derivation using formula \eqref{average n} for  the average number of rungs. The idea is that we can divide the long ladder by a rung instead of a box. In other words, if we connect two OTOCs by a rung, we will get a sum of longer ladders with certain multiplicities
\begin{equation}\label{alt_derivation}
\begin{aligned}
&
\underbrace{
\left(
\begin{tikzpicture}[baseline={([yshift=-4pt]current bounding box.center)}]
\draw[thick] (-10pt,15pt)-- (10pt,15pt);
\draw[thick] (-10pt,-15pt)-- (10pt,-15pt);
\end{tikzpicture} 
+ 
\begin{tikzpicture}[baseline={([yshift=-4pt]current bounding box.center)}]
\draw[thick, decorate, decoration={snake, segment length=11pt, amplitude=2pt
}] (0pt,-15pt) -- (0pt,15pt);
\draw[thick] (-10pt,15pt)-- (10pt,15pt);
\draw[thick] (-10pt,-15pt)-- (10pt,-15pt);
\end{tikzpicture}
+
\begin{tikzpicture}[baseline={([yshift=-4pt]current bounding box.center)}]
\draw[thick, decorate, decoration={snake, segment length=11pt, amplitude=2pt
}] (0pt,-15pt) -- (0pt,15pt);
\draw[thick, decorate, decoration={snake, segment length=11pt, amplitude=2pt
}] (20pt,-15pt) -- (20pt,15pt);
\draw[thick] (-10pt,15pt)-- (30pt,15pt);
\draw[thick] (-10pt,-15pt)-- (30pt,-15pt);
\end{tikzpicture}
+ \ldots
\right) }_{\OTOC^{\R}}
\cdot
~~~
\begin{tikzpicture}[baseline={([yshift=-4pt]current bounding box.center)}]
\draw[blue, thick, decorate, decoration={snake, segment length=11pt, amplitude=2pt
}] (0pt,-15pt) -- (0pt,15pt);
\end{tikzpicture}
~~~
\cdot 
\underbrace{
\left(
\begin{tikzpicture}[baseline={([yshift=-4pt]current bounding box.center)}]
\draw[thick] (-10pt,15pt)-- (10pt,15pt);
\draw[thick] (-10pt,-15pt)-- (10pt,-15pt);
\end{tikzpicture} 
+ 
\begin{tikzpicture}[baseline={([yshift=-4pt]current bounding box.center)}]
\draw[thick, decorate, decoration={snake, segment length=11pt, amplitude=2pt
}] (0pt,-15pt) -- (0pt,15pt);
\draw[thick] (-10pt,15pt)-- (10pt,15pt);
\draw[thick] (-10pt,-15pt)-- (10pt,-15pt);
\end{tikzpicture}
+
\begin{tikzpicture}[baseline={([yshift=-4pt]current bounding box.center)}]
\draw[thick, decorate, decoration={snake, segment length=11pt, amplitude=2pt
}] (0pt,-15pt) -- (0pt,15pt);
\draw[thick, decorate, decoration={snake, segment length=11pt, amplitude=2pt
}] (20pt,-15pt) -- (20pt,15pt);
\draw[thick] (-10pt,15pt)-- (30pt,15pt);
\draw[thick] (-10pt,-15pt)-- (30pt,-15pt);
\end{tikzpicture}
+ \ldots
\right)}_{\OTOC}
\\
&= \quad
\underbrace{
\begin{tikzpicture}[baseline={([yshift=-4pt]current bounding box.center)}]
\draw[blue, thick, decorate, decoration={snake, segment length=11pt, amplitude=2pt
}] (0pt,-15pt) -- (0pt,15pt);
\draw[thick] (-10pt,15pt)-- (10pt,15pt);
\draw[thick] (-10pt,-15pt)-- (10pt,-15pt);
\end{tikzpicture}}_{F_1}
+
\underbrace{
\begin{tikzpicture}[baseline={([yshift=-4pt]current bounding box.center)}]
\draw[blue,thick, decorate, decoration={snake, segment length=11pt, amplitude=2pt
}] (0pt,-15pt) -- (0pt,15pt);
\draw[thick, decorate, decoration={snake, segment length=11pt, amplitude=2pt
}] (20pt,-15pt) -- (20pt,15pt);
\draw[thick] (-10pt,15pt)-- (30pt,15pt);
\draw[thick] (-10pt,-15pt)-- (30pt,-15pt);
\end{tikzpicture}
+
\begin{tikzpicture}[baseline={([yshift=-4pt]current bounding box.center)}]
\draw[thick, decorate, decoration={snake, segment length=11pt, amplitude=2pt
}] (0pt,-15pt) -- (0pt,15pt);
\draw[blue,thick, decorate, decoration={snake, segment length=11pt, amplitude=2pt
}] (20pt,-15pt) -- (20pt,15pt);
\draw[thick] (-10pt,15pt)-- (30pt,15pt);
\draw[thick] (-10pt,-15pt)-- (30pt,-15pt);
\end{tikzpicture}}_{2F_2}
+ 
\underbrace{
\begin{tikzpicture}[baseline={([yshift=-4pt]current bounding box.center)}]
\draw[blue, thick, decorate, decoration={snake, segment length=11pt, amplitude=2pt
}] (0pt,-15pt) -- (0pt,15pt);
\draw[thick, decorate, decoration={snake, segment length=11pt, amplitude=2pt
}] (20pt,-15pt) -- (20pt,15pt);
\draw[thick, decorate, decoration={snake, segment length=11pt, amplitude=2pt
}] (40pt,-15pt) -- (40pt,15pt);
\draw[thick] (-10pt,15pt)-- (50pt,15pt);
\draw[thick] (-10pt,-15pt)-- (50pt,-15pt);
\end{tikzpicture}
+
\begin{tikzpicture}[baseline={([yshift=-4pt]current bounding box.center)}]
\draw[thick, decorate, decoration={snake, segment length=11pt, amplitude=2pt
}] (0pt,-15pt) -- (0pt,15pt);
\draw[blue,thick, decorate, decoration={snake, segment length=11pt, amplitude=2pt
}] (20pt,-15pt) -- (20pt,15pt);
\draw[thick, decorate, decoration={snake, segment length=11pt, amplitude=2pt
}] (40pt,-15pt) -- (40pt,15pt);
\draw[thick] (-10pt,15pt)-- (50pt,15pt);
\draw[thick] (-10pt,-15pt)-- (50pt,-15pt);
\end{tikzpicture}
+
\begin{tikzpicture}[baseline={([yshift=-4pt]current bounding box.center)}]
\draw[thick, decorate, decoration={snake, segment length=11pt, amplitude=2pt
}] (0pt,-15pt) -- (0pt,15pt);
\draw[thick, decorate, decoration={snake, segment length=11pt, amplitude=2pt
}] (20pt,-15pt) -- (20pt,15pt);
\draw[blue, thick, decorate, decoration={snake, segment length=11pt, amplitude=2pt
}] (40pt,-15pt) -- (40pt,15pt);
\draw[thick] (-10pt,15pt)-- (50pt,15pt);
\draw[thick] (-10pt,-15pt)-- (50pt,-15pt);
\end{tikzpicture}
}_{3 F_3}
\ldots
\end{aligned}
\end{equation}
The l.h.s.\ of the equation consists of one retarded OTOC (as explained in figure~\ref{fig: diag eqn}; see also Ref.~\cite{gu2019relation} section 5), a rung function, and an ordinary OTOC. The ``dot'' product represents the integration over intermediate times. There should be an additional $N$ factor on the l.h.s.\ due to the summation over internal indices. The r.h.s.\ of the equation is a sum over $n$-rung diagrams $F_n$ with a counting factor $n$. That is to say,
\begin{equation}
N \int dt_5 dt_6 
\OTOC^{\R}(t_1,t_2,t_5,t_6) \cdot R(t_{56}) \cdot \OTOC(t_5,t_6,t_3,t_4) \approx
\sum_{n=0}^{\infty} n F_n (t_1,t_2,t_3,t_4) \,,
\label{eqn: consistency}
\end{equation}
where the retarded $\OTOC^{\R}$ differs from the ordinary one by the factor of $2\cos \frac{\varkappa \pi}{2}$, namely
\begin{equation}
\OTOC^{\R}(t_1,t_2,t_3,t_4)  \approx \frac{2 \cos \frac{\varkappa \pi}{2}  }{C} 
e^{\varkappa (t_1+t_2-t_3-t_4)/2}
\VF^\R(t_{12})    \VF^\A (t_{34}) \,.
\end{equation}
\footnotetext{Here we follow the convention in Ref.~\cite{gu2019relation} (consistent with the right-to-left operator multiplication order, see e.g.\ \eqref{eqn: consistency}). One can equivalently use the more standard notation as in appendix~\ref{appendix: kelydish}, where the time goes right.}
Next, we insert the single mode ansatz into the composition formula \eqref{eqn: consistency}. 
Note that the r.h.s.\ of \eqref{eqn: consistency} is the average number of rungs times the OTOC, as discussed in section~\ref{sec: branching}. Together with \eqref{average n}, we get the following expression:
\begin{equation}\label{prove1}
\text{r.h.s.} \approx\frac{t_1+t_2-t_3-t_4}{2t_B}\OTOC(t_1,t_2,t_3,t_4)
\approx \frac{t_1+t_2-t_3-t_4}{2t_B}\frac{e^{\varkappa (t_1+t_2-t_3-t_4)/2   }}{{C}} \VF^\R(t_{12}) \VF^\A (t_{34}) \,.
\end{equation}
Meanwhile, the l.h.s.\ of \eqref{eqn: consistency} consists of two OTOCs,
\begin{equation}\label{prove2}
\text{l.h.s.}\approx \frac{2N\cos(\frac{\varkappa \pi}{2})}{C^2}\VF^R(t_{12})\VF^A(t_{34}) e^{\varkappa(t_1+t_2-t_3-t_4)/2}\underbrace{\int dt_{5}dt_6 R(t_{56})\VF^R(t_{56})\VF^A(t_{56})}_{=\frac{t_1+t_2-t_3-t_4}{2} \bigl(\VF^\A,\VF^\R\bigr)} \,,
\end{equation}
where the last integral is done by switching to new integration variables, $t_{56}=t_5-t_6$ and $t=\frac{t_5+t_6}{2}$. The latter is constrained by the end points, i.e. $\frac{t_1+t_2}{2} > t > \frac{t_3+t_4}{2}$. The integration over $t_{56}$ gives the normalization factor $\bigl(\VF^\A,\VF^\R\bigr)$ by definition, while the integration over $t$ gives the factor $\frac{t_1+t_2-t_3-t_4}{2}$. Comparing \eqref{prove1} and \eqref{prove2}, we prove the ladder identity \eqref{eqn: ladder identity}.

\section{Proof of the Lemma}
\label{appendix: proof}

In this appendix, we present a proof of the Lemma stated in the main text. The lemma asserts a bound on the rate of phase winding for the retarded and advanced fermionic Green functions. For convenience, let us denote 
\begin{equation}
I_s(\omega) : =-\frac{i}{2}  \partial_\omega \ln \frac{\tG^\R(\omega+ i s )}{\tG^\A(\omega-i s)}  \,, \quad s>0\,.
\end{equation}
The lemma asserts that 
\begin{equation}
I_s(\omega) \leq \frac{1}{s} \,, \quad \forall ~\omega \in \RR \,.
\end{equation}
Our proof relies on the positivity of the fermionic spectral function, namely, 
\begin{equation}
\tG^\R(z)=\int_{-\infty}^{+\infty} \frac{A(\omega)}{z-\omega+i\epsilon} \frac{d\omega}{2\pi} \qquad \text{with} \quad A(\omega)\geq 0\quad  \text{and}\ \int_{-\infty}^{+\infty}  A(\omega)\frac{d\omega}{2\pi} =1  \,. \label{GRspectral}
\end{equation}
The retarded Green function \eqref{GRspectral} is holomorphic in the upper half-plane. Moreover, for $\Im z > 0$, we have
\begin{equation}
-\Im \tG^\R(z) =-\Im \int_{-\infty}^{+\infty} \frac{A(\omega)}{z-\omega} \frac{d\omega}{2\pi} = \int_{-\infty}^{+\infty} \frac{A(\omega)\Im z}{|z-\omega|^2} \frac{d\omega}{2\pi}  >0 \,.
\label{eqn: imG}
\end{equation}
In other words, $f(z)=-\tG^\R(z)$ is an analytic function that maps the upper half-plane to itself. Our goal is to find a bound on the derivative of $f(z)$ at $z_0=\omega+i s$. 

To apply the basic version of the Schwarz lemma (about a bounded holomorphic map of the unit disk preserving the origin), we need to construct two maps transforming 
the upper half-plane to the unit disk and vice versa:
\begin{enumerate}
\item $y=y(\eta)$ maps the upper half-plane to the unit disk with $y(-G^\R(\omega+is))=0$, e.g.
\begin{equation}
y(\eta) = \frac{\eta+\tG^\R(\omega+is)}{\eta+(\tG^\R(\omega+is))^*} \,.
\end{equation}
We have
$
y'(-\tG^\R(\omega+is))= \frac{i}{2\Im \tG^\R(\omega+is)}$. 
\item $z=z(\xi)$ maps the unit disk to the upper half plane with $z(0)=\omega+is$. Such a map can be defined by the formula
\begin{equation}
z(\xi) = \omega + i s \cdot \frac{1-\xi}{1+\xi}
\end{equation}
and has the property $z'(0)= -2 is$. 
\end{enumerate}
Composing these two maps together with $f(z)=-\tG^\R(z)$ in the middle, we obtain a holomorphic function that maps the unit disk to itself, 
\begin{equation}
g(\xi) = y(f(z(\xi))) =\frac{-\tG^\R\left(\omega+ is \frac{1-\xi}{1+\xi}\right)+\tG^\R(\omega+is)}{-\tG^\R\left(\omega+ is \frac{1-\xi}{1+\xi}\right)+(\tG^\R(\omega+is))^*} \,,
\end{equation}
with $g(0)=0$. Therefore, according to the Schwarz lemma, we have $|g'(0)|\leq 1$, which implies
\begin{equation}
|g'(0) |=  |y'(-\tG^\R(\omega+is)) | |{\partial_\omega\tG^\R}(\omega+is)| |z'(0)| 
= \frac{s |{\partial_\omega\tG^\R}(\omega+is)| }{|\Im \tG^\R(\omega+is)|} \leq 1 \,.
\label{eqn: schwarz}
\end{equation}
Now we use the above inequality to bound $I_s(\omega)$. Let us begin with this chain of inequalities:
\begin{equation}
\begin{aligned}
I_s(\omega) &= \Im \partial_\omega \ln \tG^\R(\omega+is) \leq   | \partial_\omega \ln \tG^\R(\omega+is)| =  \left| \frac{{\partial_\omega\tG^\R}(\omega+is)}{\tG^\R(\omega+is)} \right| 
 \leq  \frac{ |{\partial_\omega\tG^\R}(\omega+is)| }{|\Im \tG^\R(\omega+is)|} \,.
\end{aligned} 
\end{equation}
Combining it with \eqref{eqn: schwarz}, we conclude that
\begin{equation}
I_s(\omega) \leq \frac{1}{s}\,.
\end{equation}

\section{Keldysh formalism for multiple contour folds}
\label{appendix: kelydish}

\begin{figure}[t]
\center
\begin{tikzpicture}[scale=0.8,baseline={([yshift=0pt]current bounding box.center)}]
\draw [->,>=stealth] (-50pt,0pt) -- (200pt,0pt) node[right]{  $\Re(t)$};
\draw [->,>=stealth] (0pt, -130pt) -- (0pt,60pt) node[right]{ $\Im(t)$};
\draw[thick,blue,far arrow] (0pt,0pt)--(140pt,0pt);
\draw[thick,blue,far arrow] (140pt,-4pt)--(0pt,-4pt);
\draw[thick,blue] (0pt,-4pt)--(0pt,-60pt);
\filldraw (140pt,-2pt) circle (2pt) node[below right]{\scriptsize contour fold $1$};
\node at (100pt,5pt) {\scriptsize $u$};
\node at (100pt,-9pt) {\scriptsize $d$};
\node at (100pt,-55pt) {\scriptsize $u$};
\node at (100pt,-69pt) {\scriptsize $d$};
\draw[thick,blue,far arrow] (0pt,-60pt)--(140pt,-60pt);
\draw[thick,blue,far arrow] (140pt,-64pt)--(0pt,-64pt);
\draw[thick,blue] (0pt,-64pt)--(0pt,-120pt);
\filldraw (140pt,-62pt) circle (2pt) node[below right]{\scriptsize contour fold $2$};
\filldraw (0pt,-120pt) circle (1pt) node[left]{\scriptsize $-i\beta$};
\filldraw (0pt,-60pt) circle (1pt) node[left]{\scriptsize $-i\tau$};
\filldraw (0pt,0pt) circle (1pt) node[below left]{\scriptsize $0$};
\end{tikzpicture}
\caption{Keldysh contour with multiple folds. Each fold has two sides, $u$ (forward time evolution) and $d$ (backward evolution). Here we use the convention (as is customary in the literature) that time runs from left to right. Note that this is different from the convention for a similar figure in Ref.~\cite{gu2019relation}, where the forward time evolution is chosen to run from right to left for the convenience of operator interpretation.}
\label{fig: OTO config}
\end{figure}
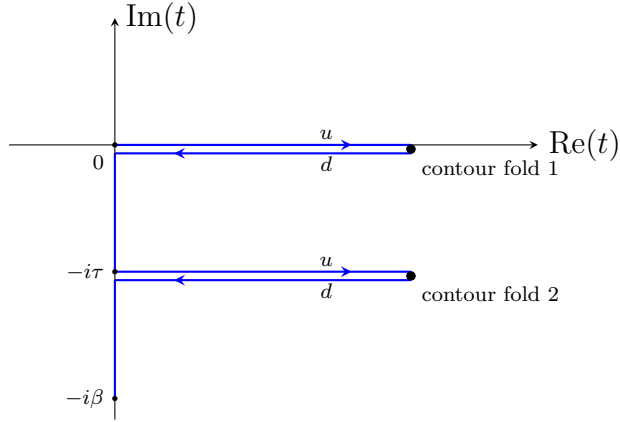

In this appendix, we study correlation functions on contours with multiple folds, as illustrated by Fig.~\ref{fig: OTO config}. In some parts, we will use the SYK model as an example. Our principal goal is to introduce the necessary language for the proof of Hermiticity of the rung function in appendix~\ref{appendix: Rung}. The multi-fold Keldysh formalism has been discussed in the literature, e.g.\ in Ref.~\cite{Aleiner:2016eni}.

The main difference from the standard Keldysh formalism is that the Green functions $G^\R$, $G^\A$, $G^\K$ are matrices with respect to the fold index ($G^\R$ and $G^\A$ are actually diagonal),  but the equations have the same form as for scalar functions. We first derive the Keldysh equations, i.e.\ a variant of the Schwinger-Dyson equation and an expression for self-energy. 

\subsection{Green function in the Keldysh basis}

Points on the contour folds are specified by the real part of time, $t$, fold index ($\alpha=1,2$), and the choice of particular side of the fold ($u$ or $d$). Fields as such are functions of complex time, so for each $t$, we have $\psi_1(t)=\psi(t)$ and $\psi_2(t)=\psi(t-i\tau)$ on the first and second fold, respectively. The distinction between $u$ and $d$ comes into play in the construction of correlation functions, which are contour-ordered. Thus, the Green function is a matrix in the $(u,d)$ basis (apart from its dependence on the fold indices):
\begin{equation}
\begin{pmatrix}
G^{uu}_{\alpha\beta} (t_1,t_2) & G^{ud}_{\alpha\beta} (t_1,t_2) \\
G^{du}_{\alpha\beta} (t_1,t_2) & G^{dd}_{\alpha\beta} (t_1,t_2)
\end{pmatrix} \,,
\qquad \text{where} \quad G^{ab}_{\alpha\beta} (t_1,t_2)
= -i 
\langle \Tc \psi^a_\alpha(t_1) {\psi^b_\beta}(t_2)^\dag  \rangle \,.
\end{equation}
The symbol $\Tc$ denotes the contour ordering: the operators are arranged such that their right-to-left order agrees with their sequence along the contour, as indicated by arrows in Fig.~\ref{fig: OTO config}. For example, $\Tc \psi^u_1(t_1) \psi^d_1(t_2)^\dag = \zeta \psi_1(t_2)^\dag\psi_1(t_1)$, where $\zeta$ is $1$ for bosons and $-1$ for fermions.

It is often convenient to introduce the $(+,-)$ basis,
\begin{equation}
|+ \rangle = \frac{| u \rangle + | d \rangle }{\sqrt{2}} \,, \quad |- \rangle = \frac{| u \rangle - | d \rangle }{\sqrt{2}}\,.
\label{normalization}
\end{equation}
Note that the $--$ correlator always vanishes. The $+-$ and $-+$ correlators are the retarded and advanced Green functions, namely, $G^{\R}=G^{+-}$ and $G^{\A}=G^{-+}$; they are diagonal in the contour index. The $++$ correlator is the Keldysh Green function, $G^{\K}=G^{++}$. We may summarize this notation in a matrix form,
\begin{equation}
\begin{pmatrix}
G^\K & G^\R \\
G^\A & 0 
\end{pmatrix} = \frac{1}{2} 
\begin{pmatrix}
1 & 1 \\
1 & -1
\end{pmatrix}
\begin{pmatrix}
G^{uu} & G^{ud} \\
G^{du} & G^{dd}
\end{pmatrix}
\begin{pmatrix}
1 & 1 \\
1 & -1
\end{pmatrix} \,.
\label{eqn: G basis}
\end{equation}
Written in terms of the field operators $\psi_{\alpha}(t)$, the Green functions are as follows:
\begin{align}
&\begin{aligned}
i G^{\R}_{\alpha \beta}(t_1,t_2) &= \delta_{\alpha \beta} \theta(t_1-t_2)\, \langle \psi_{\alpha}(t_1)\psi^\dagger_{\beta}(t_2) - \zeta \psi^\dagger_{\beta}(t_2)\psi_{\alpha }(t_1) \rangle \,, \\
i G^{\A}_{\alpha \beta}(t_1,t_2) &=- \delta_{\alpha \beta} \theta(t_2-t_1)\, \langle \psi_{\alpha}(t_1)\psi^\dagger_{\beta}(t_2) - \zeta \psi^\dagger_{\beta}(t_2)\psi_{\alpha }(t_1) \rangle \,,
\end{aligned}
\\[3pt]
&i G^{\K}_{\alpha\beta}(t_1,t_2) =
\begin{cases}
\langle \psi_\alpha (t_1) \psi_{\beta}^\dagger(t_2) + \zeta \psi_{\beta}^\dagger(t_2)\psi_{\alpha}(t_1 ) \rangle &
\text{if }\alpha =\beta\,, \\
2\zeta \langle  \psi_{\beta}^\dagger(t_2)\psi_{\alpha}(t_1 ) \rangle  &
\text{if }\alpha < \beta\,,  \\
2\langle \psi_\alpha (t_1) \psi_{\beta}^\dagger(t_2) \rangle &
\text{if }\alpha > \beta\,. 
\end{cases}
\end{align}

Let us give explicit expressions for the off-diagonal Keldysh functions at thermal equilibrium:
\begin{equation}
\begin{aligned}
i\tilde G^{\K}_{12} (\omega)
&= 2\zeta \int \langle \psi^\dagger (-i\tau) \psi(t) \rangle e^{i\omega t} dt
= \frac{2\zeta e^{(\tau-\beta)\omega}}{1-\zeta e^{-\beta \omega}} A(\omega)\,,
\\
i\tilde G^{\K}_{21} (\omega)
&= 2 \int \langle \psi (t-i\tau) \psi^\dagger(0) \rangle e^{i\omega t} dt
= \frac{2e^{-\tau\omega}}{1-\zeta e^{-\beta \omega}} A(\omega)\,,
\end{aligned}
\end{equation}
where  $A(\omega)$ is the spectral function,
\begin{equation}
A(\omega) = - 2 \Im \tG^{\R}(\omega)_{\text{equilibrium}} = i \left( \tG^{\R}(\omega) - \tG^{\A} (\omega) \right)_{\text{equilibrium}}\,.
\end{equation} 
In this notation, the Wightman function $G^\W$ we used in the main text is related to $G^\K$ for the contour with fold separation $\tau=\beta/2$ (or $\tau=\pi$ in dimensionless units):
\begin{equation}
G^\W(t) = \frac{i}{2} G^\K_{12}(t) = - \frac{i}{2} G^\K_{21}(t)\qquad
\text{for }\, \tau=\frac{\beta}{2}   \,.
\end{equation}

\subsection{Schwinger-Dyson equations}
The Keldysh equation is the Schwinger-Dyson equation,
\begin{equation}
(G_0^{-1}-\Sigma) G = 1 = G(G_0^{-1}-\Sigma)\,,
\end{equation}
where the self-energy $\Sigma$ and the inverse Green function $G_0^{-1}$ are matrices:
\begin{equation}
\Sigma = \begin{pmatrix}
0 & \Sigma^{\A} \\ 
\Sigma^{\R} & \Sigma^{\K}
\end{pmatrix} \,, \quad 
G_0^{-1} = \begin{pmatrix}
0 & \hat{\omega}  \\
\hat{\omega}  & 0
\end{pmatrix}\,.
\end{equation}
Here, $\hat{\omega}=i\partial_t$ is understood as the operator representing frequency. In terms of components, we have
\begin{gather}
G^{\R} = (\hat{\omega}- \Sigma^{\R})^{-1}\,,\qquad
G^{\A} = (\hat{\omega} - \Sigma^{\A})^{-1} \,,\\[2pt]
(\hat{\omega} -\Sigma^\R) G^\K - \Sigma^\K G^\A = 0= G^\K(\hat{\omega}-\Sigma^A) - G^\R \Sigma^\K\,. \label{eqn: keldysh}
\end{gather}

\subsection{Diagrammatic rules}
Now, we derive the diagrammatic rules using the SYK model as an example. In addition to the Green functions, we will need the interaction term on the Keldysh contour:
\begin{equation}\label{eqn:int}
\begin{aligned}
\calL_\text{int}=&-\sum_{j_1,...,j_q,\alpha}i^{\frac{q}{2}}\frac{J_{j_1,...,j_q}}{q!}\left(\psi^u_{\alpha,j_1}\psi^u_{\alpha,j_2}...\psi^u_{\alpha,j_q}-\psi^d_{\alpha,j_1}\psi^d_{\alpha,j_2}...\psi^d_{\alpha,j_q}\right)\\
=&-\sum_{j_1,...,j_q,\alpha}i^{\frac{q}{2}}\frac{J_{j_1,...,j_q}}{q!}\left(\sum_{k~\text{odd}}2^{1-\frac{q}{2}} {q \choose k}\psi^+_{\alpha,j_1}\psi^+_{\alpha,j_2}...\psi^+_{\alpha,j_k}\psi^-_{\alpha,j_{k+1}}...\psi^-_{\alpha,j_q}\right)\,.
\end{aligned}
\end{equation}
The prefactor $2^{1-q/2}=\bigl(2^{-1/2}\bigr)^q\cdot 2$ in the second line arises as follows: $2^{-1/2}$ comes from the basis change formula \eqref{normalization}, and the additional factor of $2$ is due to duplicate contributions from the two terms in the first line. The factor ${q \choose k}$ counts different choices leading to the same term with $k$ ``$+$'' fields.

We are now in a position to formulate the Feynman rules specific to the model.
\begin{enumerate}
\item Green function in $(+,-)$ basis:
\begin{equation}
\begin{tikzpicture}[baseline={([yshift=0pt]current bounding box.center)}]
\draw[thick, mid arrow] (60pt,0pt)--(0pt,0pt) ;
\filldraw (0pt,0pt) circle (1pt) node[below]{$t_1,\alpha$};
\filldraw (60pt,0pt) circle (1pt) node[below]{$t_2,\beta$};
\node[above] at (0pt,0pt) {$(+,-)$};
\node[above] at (60pt,0pt) {$(+,-)$};
\end{tikzpicture}
 \qquad 
=\qquad 
  i \begin{pmatrix}
G_{\alpha\beta}^\K(t_1,t_2) & G_{\alpha\beta}^\R(t_1,t_2) \\
G_{\alpha\beta}^\A(t_1,t_2) & 0 
\end{pmatrix}\,.
\label{GreenDef}
\end{equation}
Note the factor of $i$ in the definition. 
\item Interaction  vertex in the $(u,d)$ basis (for the purpose of illustration, we draw the diagrams for $q=4$):
\begin{equation}
\begin{tikzpicture}[baseline={([yshift=-4pt]current bounding box.center)}]
\filldraw (0pt,0pt) circle (1pt);
\draw[thick] (0pt,0pt) -- (-15pt,15pt) node[left]{$u,\alpha$};
\draw[thick] (0pt,0pt) -- (15pt,15pt) node[right]{$u,\alpha$};
\draw[thick] (0pt,0pt) -- (-15pt,-15pt) node[left]{$u,\alpha$};
\draw[thick] (0pt,0pt) -- (15pt,-15pt) node[right]{$u,\alpha$};
\end{tikzpicture} \quad=\quad i^{\frac{q}{2}-1} J_{jklm}\,, \qquad\quad 
\begin{tikzpicture}[baseline={([yshift=-4pt]current bounding box.center)}]
\filldraw (0pt,0pt) circle (1pt);
\draw[thick] (0pt,0pt) -- (-15pt,15pt) node[left]{$d,\alpha$};
\draw[thick] (0pt,0pt) -- (15pt,15pt) node[right]{$d,\alpha$};
\draw[thick] (0pt,0pt) -- (-15pt,-15pt) node[left]{$d,\alpha$};
\draw[thick] (0pt,0pt) -- (15pt,-15pt) node[right]{$d,\alpha$};
\end{tikzpicture} \quad\,=\quad\, -i^{\frac{q}{2}-1}  J_{jklm}\,.
\label{VertexDef}
\end{equation}
In the $(+,-)$ basis,
\begin{equation}
\begin{tikzpicture}[baseline={([yshift=-4pt]current bounding box.center)}]
\filldraw (0pt,0pt) circle (1pt);
\draw[thick] (0pt,0pt) -- (-15pt,15pt) node[left]{$-,\alpha$};
\draw[thick] (0pt,0pt) -- (15pt,15pt) node[right]{$+,\alpha$};
\draw[thick] (0pt,0pt) -- (-15pt,-15pt) node[left]{$+,\alpha$};
\draw[thick] (0pt,0pt) -- (15pt,-15pt) node[right]{$+,\alpha$};
\end{tikzpicture} \quad\,=\quad\,
\begin{tikzpicture}[baseline={([yshift=-4pt]current bounding box.center)}]
\filldraw (0pt,0pt) circle (1pt);
\draw[thick] (0pt,0pt) -- (-15pt,15pt) node[left]{$+,\alpha$};
\draw[thick] (0pt,0pt) -- (15pt,15pt) node[right]{$-,\alpha$};
\draw[thick] (0pt,0pt) -- (-15pt,-15pt) node[left]{$-,\alpha$};
\draw[thick] (0pt,0pt) -- (15pt,-15pt) node[right]{$-,\alpha$};
\end{tikzpicture} \quad\,=\quad\,
i^{\frac{q}{2}-1}  2^{1-\frac{q}{2}}J_{jklm}\,.
\label{pmvertex}
\end{equation}
\end{enumerate}

The construction of diagrams involves one more step: Gaussian averaging over $J_{jklm}$ is represented by dotted lines connecting pairs of vertices.

\subsection{Expression for the self-energy}
Neglecting subleading (in $1/N$) terms, we have
\begin{equation}
-i
\Sigma_{\alpha \beta}(t_1,t_2)\,\,=\,\frac{1}{(q-1)!}\,\,
\begin{tikzpicture}[baseline={([yshift=-7pt]current bounding box.center)}]
\filldraw (-20pt,0pt) circle (1pt);
\filldraw (20pt,0pt) circle (1pt);
\draw[thick] (-20pt,0pt) .. controls (-8pt,-12pt) and (8pt,-12pt) .. (20pt,0pt);
\draw[thick] (20pt,0pt) .. controls (8pt,12pt) and (-8pt,12pt) .. (-20pt,0pt);
\draw[thick] (20pt,0pt) -- (-20pt,0pt);
\draw[thick] (-20pt,0pt) -- (-40pt,0pt);
\draw[thick] (20pt,0pt) -- (40pt,0pt);
\node at (-30pt,4pt) {\scriptsize $t_1,\alpha$};
\node at (30pt,4pt) {\scriptsize $t_2,\alpha$};
\draw[thick, dotted] (20pt,0pt) .. controls (8pt,20pt) and (-8pt,20pt) .. (-20pt,0pt);
\end{tikzpicture}\,\,,
\label{SelfEnergyDef}
\end{equation}
where $(q-1)!$ in the denominator is the number of symmetries and the dotted line represents the disorder averaging, which gives a factor $J^2(q-1)!$. Both sides of the above equation are matrices in the $(u,d)$ or $(+,-)$ basis. For the calculation of its elements, we should sum over patterns of pluses and minuses with an odd number of pluses around each vertex. Not all such patterns are allowed because $G^{--}=0$ and because the product of the retarded and advanced Green functions vanishes, i.e.\ $G^{+-}_{\alpha\beta}(t_1,t_2) G^{-+}_{\alpha\beta}(t_1,t_2)=0$. In particular, the allowed Green functions in the expression for $\Sigma^{\R}=\Sigma^{-+}$ are $G^{\R}=G^{+-}$ and $G^{\K}=G^{++}$:
\begin{equation}
\begin{tikzpicture}[baseline={([yshift=-4pt]current bounding box.center)}]
\filldraw (-20pt,0pt) circle (1pt);
\filldraw (20pt,0pt) circle (1pt);
\draw[thick] (-20pt,0pt) .. controls (-8pt,-12pt) and (8pt,-12pt) .. (20pt,0pt);
\draw[thick] (20pt,0pt) .. controls (8pt,12pt) and (-8pt,12pt) .. (-20pt,0pt);
\draw[thick] (20pt,0pt) -- (-20pt,0pt);
\draw[thick] (-20pt,0pt) -- (-35pt,0pt);
\draw[thick] (20pt,0pt) -- (35pt,0pt);
\node at (-30pt,4pt) {\scriptsize $-$};
\node at (30pt,4pt) {\scriptsize $+$};
\draw[thick, dotted] (20pt,0pt) .. controls (8pt,20pt) and (-8pt,20pt) .. (-20pt,0pt);
\end{tikzpicture} 
\quad\,=\quad\,
\begin{tikzpicture}[baseline={([yshift=-4pt]current bounding box.center)}]
\filldraw (-20pt,0pt) circle (1pt);
\filldraw (20pt,0pt) circle (1pt);
\draw[thick] (-20pt,0pt) .. controls (-8pt,-12pt) and (8pt,-12pt) .. (20pt,0pt);
\draw[thick] (20pt,0pt) .. controls (8pt,12pt) and (-8pt,12pt) .. (-20pt,0pt);
\draw[thick] (20pt,0pt) -- (-20pt,0pt);
\draw[thick] (-20pt,0pt) -- (-35pt,0pt);
\draw[thick] (20pt,0pt) -- (35pt,0pt);
\node at (-30pt,4pt) {\scriptsize $-$};
\node at (30pt,4pt) {\scriptsize $+$};
\node at (-15pt,10pt) {\scriptsize $+$};
\node at (-15pt,-10pt) {\scriptsize $+$};
\node at (15pt,10pt) {\scriptsize $-$};
\node at (15pt,-10pt) {\scriptsize $-$};
\node at (-6pt,4pt){\scriptsize $+$};
\node at (6pt,4pt){\scriptsize $-$};
\draw[thick, dotted] (20pt,0pt) .. controls (8pt,20pt) and (-8pt,20pt) .. (-20pt,0pt);
\end{tikzpicture} 
\,\,+\,\,3\,\,
\begin{tikzpicture}[baseline={([yshift=-4pt]current bounding box.center)}]
\filldraw (-20pt,0pt) circle (1pt);
\filldraw (20pt,0pt) circle (1pt);
\draw[thick] (-20pt,0pt) .. controls (-8pt,-12pt) and (8pt,-12pt) .. (20pt,0pt);
\draw[thick] (20pt,0pt) .. controls (8pt,12pt) and (-8pt,12pt) .. (-20pt,0pt);
\draw[thick] (20pt,0pt) -- (-20pt,0pt);
\draw[thick] (-20pt,0pt) -- (-35pt,0pt);
\draw[thick] (20pt,0pt) -- (35pt,0pt);
\node at (-30pt,4pt) {\scriptsize $-$};
\node at (30pt,4pt) {\scriptsize $+$};
\node at (-15pt,10pt) {\scriptsize $+$};
\node at (-15pt,-10pt) {\scriptsize $+$};
\node at (15pt,10pt) {\scriptsize $-$};
\node at (15pt,-10pt) {\scriptsize $+$};
\node at (-6pt,4pt){\scriptsize $+$};
\node at (6pt,4pt){\scriptsize $+$};
\draw[thick, dotted] (20pt,0pt) .. controls (8pt,20pt) and (-8pt,20pt) .. (-20pt,0pt);
\end{tikzpicture}\,\,.
\end{equation}
In the above equation, we have assumed that $q=4$. For a general $q$, the series continues as follows:
\begin{equation}
\Sigma^\R_{\alpha \beta} (t_1,t_2)= \frac{i^{q-2} (q-1)! J^2}{2^{q-2}}
\sum_{1\leq k < q,  \text{odd}} \frac{ 1}{(q-k)!(k-1)!}  G_{\alpha\beta}^\R(t_1,t_2)^{q-k} G_{\alpha\beta}^\K(t_1,t_2)^{k-1} \,.
\label{eq: self energy}
\end{equation}
The prefactor $i^{q-2}/2^{q-2}$ comes from \eqref{pmvertex}, and $(q-1)!J^2$ is due to disorder averaging. The combinatorial factor $\frac{1}{(q-k)!(k-1)!}$ is the inverse number of symmetries of the diagram, which has $q-k$ retarded (i.e.\ $+-$) lines and $k-1$ Keldysh (i.e.\ $++$) lines. There are also some implicit factors that cancel each other:  $(-1)^{\frac{(q-1)q}{2}}$ (fermionic sign), $i^{q-1}$ from $q-1$ Green functions (see \eqref{GreenDef}), and $i$ from the definition of self-energy (see \eqref{SelfEnergyDef}). We may transform the above series into a more compact form:
\begin{equation}
\Sigma^\R_{\alpha \beta} (t_1,t_2) = 
\frac{i^{q-2}  J^2}{2^{q-1}} \left[\left( G^\R_{\alpha \beta} (t_1,t_2) +G^K_{\alpha \beta} (t_1,t_2)\right)^{q-1} +
\left( G^\R_{\alpha \beta} (t_1,t_2) -G^K_{\alpha \beta} (t_1,t_2)\right)^{q-1}  
\right]\,.
\end{equation}
Similarly, the advanced self-energy is given by this expression:
\begin{equation}
\Sigma^\A_{\alpha \beta} (t_1,t_2)= 
\frac{i^{q-2}  J^2}{2^{q-1}} \left[\left( G^\A_{\alpha \beta} (t_1,t_2) +G^K_{\alpha \beta} (t_1,t_2)\right)^{q-1} +
\left( G^\A_{\alpha \beta} (t_1,t_2) -G^K_{\alpha \beta} (t_1,t_2)\right)^{q-1}  
\right]\,.
\end{equation}
The Keldysh self-energy needs to be derived separately, 
\begin{equation}
\begin{tikzpicture}[baseline={([yshift=-4pt]current bounding box.center)}]
\filldraw (-20pt,0pt) circle (1pt);
\filldraw (20pt,0pt) circle (1pt);
\draw[thick] (-20pt,0pt) .. controls (-8pt,-12pt) and (8pt,-12pt) .. (20pt,0pt);
\draw[thick] (20pt,0pt) .. controls (8pt,12pt) and (-8pt,12pt) .. (-20pt,0pt);
\draw[thick] (20pt,0pt) -- (-20pt,0pt);
\draw[thick] (-20pt,0pt) -- (-35pt,0pt);
\draw[thick] (20pt,0pt) -- (35pt,0pt);
\node at (-30pt,4pt) {\scriptsize $-$};
\node at (30pt,4pt) {\scriptsize $-$};
\draw[thick, dotted] (20pt,0pt) .. controls (8pt,20pt) and (-8pt,20pt) .. (-20pt,0pt);
\end{tikzpicture} 
\quad\,=\quad\,
\begin{tikzpicture}[baseline={([yshift=-4pt]current bounding box.center)}]
\filldraw (-20pt,0pt) circle (1pt);
\filldraw (20pt,0pt) circle (1pt);
\draw[thick] (-20pt,0pt) .. controls (-8pt,-12pt) and (8pt,-12pt) .. (20pt,0pt);
\draw[thick] (20pt,0pt) .. controls (8pt,12pt) and (-8pt,12pt) .. (-20pt,0pt);
\draw[thick] (20pt,0pt) -- (-20pt,0pt);
\draw[thick] (-20pt,0pt) -- (-35pt,0pt);
\draw[thick] (20pt,0pt) -- (35pt,0pt);
\node at (-30pt,4pt) {\scriptsize $-$};
\node at (30pt,4pt) {\scriptsize $-$};
\node at (-15pt,10pt) {\scriptsize $+$};
\node at (-15pt,-10pt) {\scriptsize $+$};
\node at (15pt,10pt) {\scriptsize $+$};
\node at (15pt,-10pt) {\scriptsize $+$};
\node at (-6pt,4pt){\scriptsize $+$};
\node at (6pt,4pt){\scriptsize $+$};
\draw[thick, dotted] (20pt,0pt) .. controls (8pt,20pt) and (-8pt,20pt) .. (-20pt,0pt);
\end{tikzpicture} 
\,\,+\,\,3\,\,
\begin{tikzpicture}[baseline={([yshift=-4pt]current bounding box.center)}]
\filldraw (-20pt,0pt) circle (1pt);
\filldraw (20pt,0pt) circle (1pt);
\draw[thick] (-20pt,0pt) .. controls (-8pt,-12pt) and (8pt,-12pt) .. (20pt,0pt);
\draw[thick] (20pt,0pt) .. controls (8pt,12pt) and (-8pt,12pt) .. (-20pt,0pt);
\draw[thick] (20pt,0pt) -- (-20pt,0pt);
\draw[thick] (-20pt,0pt) -- (-35pt,0pt);
\draw[thick] (20pt,0pt) -- (35pt,0pt);
\node at (-30pt,4pt) {\scriptsize $-$};
\node at (30pt,4pt) {\scriptsize $-$};
\node at (-15pt,10pt) {\scriptsize $+$};
\node at (-15pt,-10pt) {\scriptsize $+$};
\node at (15pt,10pt) {\scriptsize $-$};
\node at (15pt,-10pt) {\scriptsize $+$};
\node at (-6pt,4pt){\scriptsize $+$};
\node at (6pt,4pt){\scriptsize $-$};
\draw[thick, dotted] (20pt,0pt) .. controls (8pt,20pt) and (-8pt,20pt) .. (-20pt,0pt);
\end{tikzpicture} 
\,\,+\,\,3\,\,
\begin{tikzpicture}[baseline={([yshift=-4pt]current bounding box.center)}]
\filldraw (-20pt,0pt) circle (1pt);
\filldraw (20pt,0pt) circle (1pt);
\draw[thick] (-20pt,0pt) .. controls (-8pt,-12pt) and (8pt,-12pt) .. (20pt,0pt);
\draw[thick] (20pt,0pt) .. controls (8pt,12pt) and (-8pt,12pt) .. (-20pt,0pt);
\draw[thick] (20pt,0pt) -- (-20pt,0pt);
\draw[thick] (-20pt,0pt) -- (-35pt,0pt);
\draw[thick] (20pt,0pt) -- (35pt,0pt);
\node at (-30pt,4pt) {\scriptsize $-$};
\node at (30pt,4pt) {\scriptsize $-$};
\node at (-15pt,10pt) {\scriptsize $-$};
\node at (-15pt,-10pt) {\scriptsize $+$};
\node at (15pt,10pt) {\scriptsize $+$};
\node at (15pt,-10pt) {\scriptsize $+$};
\node at (-6pt,4pt){\scriptsize $-$};
\node at (6pt,4pt){\scriptsize $+$};
\draw[thick, dotted] (20pt,0pt) .. controls (8pt,20pt) and (-8pt,20pt) .. (-20pt,0pt);
\end{tikzpicture}\,\,,
\end{equation}
\begin{equation}
\Sigma^\K_{\alpha \beta} (t_1,t_2) = \frac{i^{q-2}  J^2}{2^{q-2}} 
\Bigg[
\sum_{1\leq k < q,  \text{odd}}  {q-1 \choose k-1}  G_{\alpha\beta}^\K(t_1,t_2)^{q-k}  \left( G^\R_{\alpha\beta}(t_1,t_2)^{k-1} +G^\A_{\alpha\beta}(t_1,t_2)^{k-1} -\delta_{k,1} \right) \Bigg] 
\end{equation}
where the last term is to compensate the double counting in the summation when $k=1$. For later convenience, we may rewrite
\begin{equation}
G^\R_{\alpha\beta}(t_1,t_2)^{k-1} +G^\A_{\alpha\beta}(t_1,t_2)^{k-1}   = (G^\R_{\alpha\beta}(t_1,t_2)- G^\A_{\alpha\beta}(t_1,t_2)  )^{k-1} \,, \quad \text{(for odd $k>1$)} 
\end{equation}
since the mixed terms all vanish due to time constraints for the retarded and advanced Green functions. Again, the $k=1$ term needs individual care, and we find
\begin{equation}
\begin{aligned}
\Sigma^\K_{\alpha \beta} (t_1,t_2) = \frac{i^{q-2}  J^2}{2^{q-1}} &
\left[ \left(G_{\alpha\beta}^\K(t_1,t_2)+  G^\R_{\alpha\beta}(t_1,t_2) -G^\A_{\alpha\beta}(t_1,t_2)  \right)^{q-1} \right. \\
&\left.
+ 
\left(G_{\alpha\beta}^\K(t_1,t_2)-  G^\R_{\alpha\beta}(t_1,t_2) +G^\A_{\alpha\beta}(t_1,t_2)  \right)^{q-1}
  \right] \,.
\end{aligned}
\end{equation}
Note that the retarded and advanced Green functions $G^{\R, \A}_{\alpha\beta}$ are diagonal in the fold index (i.e.\ vanish if $\alpha\not=\beta$); therefore the corresponding self-energies $\Sigma^{\R, \A}_{\alpha\beta}$ are also diagonal. In contrast, the Keldysh component $\Sigma^\K_{\alpha\beta}$ is not necessarily diagonal. 

An alternative way to obtain the above relations is to work in the $(u,d)$ basis, where the vertex is diagonal  
\begin{equation}
\begin{pmatrix}
\Sigma^{uu} & \Sigma^{ud} \\
\Sigma^{du} & \Sigma^{dd} 
\end{pmatrix} = 
i J^2 
\begin{pmatrix}
-(iG^{uu})^{q-1} & (iG^{ud})^{q-1} \\
(iG^{du})^{q-1} & -(iG^{dd})^{q-1}
\end{pmatrix} \,.
\end{equation}
Besides the basis transformation law of $G$ shown in \eqref{eqn: G basis}, one needs to use a similar rule for $\Sigma$:
\begin{equation}
\begin{pmatrix}
0 & \Sigma^\A \\
\Sigma^\R & \Sigma^\K
\end{pmatrix} = \frac{1}{2} 
\begin{pmatrix}
1 & 1 \\
1 & -1
\end{pmatrix}
\begin{pmatrix}
\Sigma^{uu} & \Sigma^{ud} \\
\Sigma^{du} & \Sigma^{dd}
\end{pmatrix}
\begin{pmatrix}
1 & 1 \\
1 & -1
\end{pmatrix} \,.
\label{eqn: Sigma basis}
\end{equation}

\subsection{Quasiparticle decay rate $\Gamma$ for SYK at weak coupling}
\label{appendix: gamma}

\begin{figure}[t]
\center
\begin{tikzpicture}[yscale=6, xscale=0.36, baseline={(current bounding box.center)}]
\draw[->,>=stealth] (2,0.707) -- (20,0.707)  node[right]{$q$};
\draw[->,>=stealth] (2,0.707) -- (2,1.2)  node[above]{$\Gamma/\Gamma_*$};
\draw[domain=2:18,smooth,variable=\q,blue] plot ({\q},{sqrt(\q/(2*(\q-1)))});
\draw[domain=2:18,smooth,variable=\q,red] plot ({\q},{1});
\draw (2,0.998831) ellipse (5pt and 0.3pt);
\draw (4,1.0585) ellipse (5pt and 0.3pt);
\draw (6,1.04011) ellipse (5pt and 0.3pt);
\draw (8,1.02971) ellipse (5pt and 0.3pt);
\draw (10,1.02291) ellipse (5pt and 0.3pt);
\draw (12,1.0182) ellipse (5pt and 0.3pt);
\draw (14,1.01467) ellipse (5pt and 0.3pt);
\draw (16,1.0119) ellipse (5pt and 0.3pt);
\draw (18,1.00964) ellipse (5pt and 0.3pt);
\node at (2,1) [left]{\scriptsize $1$};
\node at (2,0.707) [left]{\scriptsize $1/\sqrt{2}$};
\node at (2,0.707) [below]{\scriptsize $2$};
\node at (4,0.707) [below]{\scriptsize $4$};
\node at (6,0.707) [below]{\scriptsize $6$};
\node at (8,0.707) [below]{\scriptsize $8$};
\node at (10,0.707) [below]{\scriptsize $10$};
\node at (12,0.707) [below]{\scriptsize $12$};
\node at (14,0.707) [below]{\scriptsize $14$};
\node at (16,0.707) [below]{\scriptsize $16$};
\node at (18,0.707) [below]{\scriptsize $18$};
\end{tikzpicture}
\caption{Quasiparticle decay rate $\Gamma$ in units of $\Gamma_*=\sqrt{\frac{2^{5-q}}{q}} J$ as a function of $q$. The red line, $\Gamma/\Gamma_*=1$, is the analytic large $q$ result, whereas the blue line, $\Gamma/\Gamma_*=\sqrt{\frac{q}{2(q-1)}}$, represents the estimate \eqref{eqn: estimation gamma}. The black circles are numerical results for even integer $q$ at $\beta=0$.}
\label{fig: decay rate}
\end{figure}
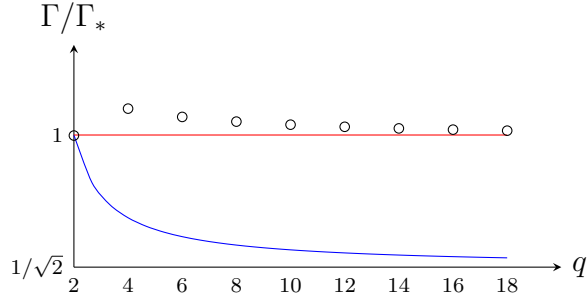

As a simple application of the formalism, we will estimate the quasiparticle decay rate $\Gamma$ for the SYK model at $\beta J\ll 1$. This quantity is defined by the $t\to\infty$ asymptotics of the Green function, $G^{\R}(t,0) \sim e^{-\Gamma t/2}$. Let us make a stronger assumption and adopt the ansatz $\tG^\R(\omega) \approx \frac{1}{\omega+i\Gamma/2}$, which is only qualitatively correct. The task is to self-consistently determine $\Gamma$ using the equations for the Green function and self-energy. First, by definition,
\begin{equation}
A(\omega) = -2 \Im \tG^\R(\omega) \approx \frac{\Gamma}{\omega^2+\Gamma^2/4}\,, \quad \tG^{\K}(\omega) = - i \frac{1-e^{-\beta \omega}}{1+e^{-\beta \omega}} A(\omega)\,.
\end{equation}
For a large $t$, we have 
\begin{equation}
G^\R(t,0) \approx -i \theta (t) e^{-\Gamma t/2}\,, \quad G^\K(t,0) \approx -i \frac{1-e^{i\Gamma \beta/2}}{1+e^{i\Gamma \beta/2}} e^{-\Gamma |t|/2} \,.
\label{eqn: GR GK example}
\end{equation}
Inserting these expressions into \eqref{eq: self energy}, we obtain an explicit formula for the retarded self-energy:
\begin{equation}\label{SigmaR_large_t}
\Sigma^\R(t,0) \approx -i \frac{J^2 }{2^{q-2}} \theta(t) e^{-(q-1)\Gamma t/2} \underbrace{\sum_{1\leq k< q, {\rm odd}}  {q-1 \choose k-1} x^{k-1}}_{= \frac{1}{2} \left[(1+x)^{q-1}+(1-x)^{q-1}\right]}\,, \quad x=\frac{1-e^{i\Gamma \beta/2}}{1+e^{i\Gamma \beta/2}}
\end{equation}
We are interested in the high temperature limit, where $x \approx -i \frac{\Gamma \beta}{4} $ is small, and therefore, negligible for the leading order calculation. (In other words, the diagonal component of the Keldysh function is small, $|G^\K|\ll |G^\R|$.) Next, we estimate the zero frequency value of the self-energy by integrating its long-time asymptotic form \eqref{SigmaR_large_t},\footnote{This is a crude approximation, and we will comment on its limitations shortly.} i.e. 
\begin{equation}
\tilde \Sigma^\R(0) \approx  -i \frac{J^2 }{(q-1)  2^{q-3}} \frac{1}{\Gamma} \,.
\label{eqn: zero freq self energy}
\end{equation}
This quantity should be set to $-i\Gamma/2$ as a self-consistency condition. Thus, we obtain 
\begin{equation}
\Gamma \approx \frac{1}{\sqrt{q-1}\, 2^{q/2-2}}\, J\,.
\label{eqn: estimation gamma}
\end{equation}
We now make a few remarks about this formula.
\begin{enumerate}
\item The quasiparticle decay rate $\Gamma$ is of order $J$ instead of $J^2$; the latter is what one might have guessed based on perturbation theory as all disorder-averaged diagrams come with even powers of $J$. However, the naive perturbative argument fails due to the divergence arising from the long-lived bare Green function. We need to include the quasiparticle decay in the first place. On the other hand, $\Gamma\sim J$ is a reasonable result from the dimensional analysis perspective since $J$ is the only energy scale at infinite temperature.

\item We should not trust the order $1$ prefactor because in \eqref{eqn: zero freq self energy}, we have used the long-time asymptotics to approximate $\Sigma^\R(t)$. It is not valid at times $t \lesssim 1/\Gamma$, and therefore, the expression for $\Gamma$ is off by an order $1$ factor, though the $J$ scaling is correct. We expect the approximation \eqref{eqn: zero freq self energy} and the final formula to become accurate in the $q\rightarrow 2$ limit. Indeed, as shown in Fig.~\ref{fig: decay rate}, the approximate formula \eqref{eqn: estimation gamma} agrees with the numerical solution for $q=2$. 

\item We can also compare \eqref{eqn: estimation gamma} with the exact large $q$ result (see Appendix~\ref{appendix: large q}), which we denote by $\Gamma_*$. Specifically, $\Gamma_{*}= 2vq^{-1}$ and $v\approx 2^{3/2-q/2} \sqrt{q} J$ for weak coupling. There is some discrepancy, with $\Gamma$ given by \eqref{eqn: estimation gamma} equal to $\Gamma_*/\sqrt{2}$ in the $q\to\infty$ limit. This is not a surprise for the reason discussed in point 2. 
\end{enumerate}

\section{Hermiticity of $R(t)$}
\label{appendix: Rung}

In this appendix, we show the Hermiticity of the rung function, namely
\begin{equation}
R_{ab}(t) = R_{ba}^*(t)\,,  \qquad\text{i.e.} \qquad 
\begin{tikzpicture}[baseline={([yshift=-4pt]current bounding box.center)}]
\draw[mid arrow, thick] (15pt,20pt)--(0pt,20pt);
\draw[mid arrow, thick] (0pt,20pt)--(-15pt,20pt);
\draw[mid arrow, thick] (-15pt,-20pt)--(0pt,-20pt);
\draw[mid arrow, thick] (0pt,-20pt)--(15pt,-20pt);
\draw[thick, decorate, decoration={snake, segment length=11pt, amplitude=2pt
}] (0pt,20pt) -- (0pt,-20pt);
\node[left] at (-15pt,20pt) {$a$};
\node[right] at (15pt,20pt) {$b$};
\node[left] at (-15pt,-20pt) {$a$};
\node[right] at (15pt,-20pt) {$b$};
\end{tikzpicture} 
=
\left(
\begin{tikzpicture}[baseline={([yshift=-4pt]current bounding box.center)}]
\draw[mid arrow, thick] (15pt,20pt)--(0pt,20pt);
\draw[mid arrow, thick] (0pt,20pt)--(-15pt,20pt);
\draw[mid arrow, thick] (-15pt,-20pt)--(0pt,-20pt);
\draw[mid arrow, thick] (0pt,-20pt)--(15pt,-20pt);
\draw[thick, decorate, decoration={snake, segment length=11pt, amplitude=2pt
}] (0pt,20pt) -- (0pt,-20pt);
\node[left] at (-15pt,20pt) {$b$};
\node[right] at (15pt,20pt) {$a$};
\node[left] at (-15pt,-20pt) {$b$};
\node[right] at (15pt,-20pt) {$a$};
\end{tikzpicture} 
\right)^* .
\end{equation}
where $a,b$ label field flavors (not to be confused with fold indices $\alpha,\beta$). Here we assume that the flavors on the top and the bottom rails come in identical pairs. That is a nontrivial condition, restricting the class of models we can handle.

To formulate the problem, we consider the interaction vertex $\lambda_{abc} \psi_a^\dagger \psi_b O_c$, where $O_c$ is a bosonic field that mediates the interaction between the rails (cf.\ Eq.~$(5.1)$ in Ref.~\cite{gu2019relation}). Without loss of generality, we choose $O_c$ to be real; namely, we treat the real and imaginary parts of a complex field separately. In the Keldysh basis $(u,d)$, the interaction is written as follows:
\begin{equation}
\begin{aligned}
&\mathcal{L}_{\rm int} = -i \lambda_{abc} \left[ (\psi^u_a)^\dagger \psi_b^u O_c^u -(\psi^d_a)^\dagger \psi_b^d O_c^d \right]
\\
 \text{i.e.} \quad&
\begin{tikzpicture}[baseline={(current bounding box.center)}]
\filldraw (0pt,0pt) circle (1pt);
\draw[thick] (-20pt,0pt)--(20pt,0pt);
\draw[thick, decorate, decoration={snake, segment length=11pt, amplitude=2pt
}] (0pt,0pt) -- (0pt,-20pt);
\node[above] at (-18pt,0pt) {$ $};
\node[above] at (18pt,0pt) {$ $};
\node[right] at (0pt,-18pt) {$ $};
\end{tikzpicture} =
\begin{tikzpicture}[baseline={(current bounding box.center)}]
\filldraw (0pt,0pt) circle (1pt);
\draw[thick] (-20pt,0pt)--(20pt,0pt);
\draw[thick, decorate, decoration={snake, segment length=11pt, amplitude=2pt
}] (0pt,0pt) -- (0pt,-20pt);
\node[above] at (-18pt,0pt) {$u$};
\node[above] at (18pt,0pt) {$u$};
\node[right] at (0pt,-18pt) {$u$};
\end{tikzpicture} 
-
\begin{tikzpicture}[baseline={(current bounding box.center)}]
\filldraw (0pt,0pt) circle (1pt);
\draw[thick] (-20pt,0pt)--(20pt,0pt);
\draw[thick, decorate, decoration={snake, segment length=11pt, amplitude=2pt
}] (0pt,0pt) -- (0pt,-20pt);
\node[above] at (-18pt,0pt) {$d$};
\node[above] at (18pt,0pt) {$d$};
\node[right] at (0pt,-18pt) {$d$};
\end{tikzpicture} 
\end{aligned}
\end{equation}
In the $(+,-)$ basis, 
\begin{equation}
\begin{tikzpicture}[scale=1.0, baseline={(current bounding box.center)}]
\filldraw (0pt,0pt) circle (1pt);
\draw[thick] (-20pt,0pt)--(20pt,0pt);
\draw[thick, decorate, decoration={snake, segment length=11pt, amplitude=2pt
}] (0pt,0pt) -- (0pt,-20pt);
\end{tikzpicture} = 
\frac{1}{\sqrt{2}} \left( 
\begin{tikzpicture}[scale=1.0, baseline={(current bounding box.center)}]
\filldraw (0pt,0pt) circle (1pt);
\draw[thick] (-20pt,0pt)--(20pt,0pt);
\draw[thick, decorate, decoration={snake, segment length=11pt, amplitude=2pt
}] (0pt,0pt) -- (0pt,-20pt);
\node[above] at (-18pt,0pt) {$+$};
\node[above] at (18pt,0pt) {$+$};
\node[right] at (0pt,-18pt) {$-$};
\end{tikzpicture}
+
\begin{tikzpicture}[scale=1.0, baseline={(current bounding box.center)}]
\filldraw (0pt,0pt) circle (1pt);
\draw[thick] (-20pt,0pt)--(20pt,0pt);
\draw[thick, decorate, decoration={snake, segment length=11pt, amplitude=2pt
}] (0pt,0pt) -- (0pt,-20pt);
\node[above] at (-18pt,0pt) {$+$};
\node[above] at (18pt,0pt) {$-$};
\node[right] at (0pt,-18pt) {$+$};
\end{tikzpicture}
+
\begin{tikzpicture}[scale=1.0, baseline={(current bounding box.center)}]
\filldraw (0pt,0pt) circle (1pt);
\draw[thick] (-20pt,0pt)--(20pt,0pt);
\draw[thick, decorate, decoration={snake, segment length=11pt, amplitude=2pt
}] (0pt,0pt) -- (0pt,-20pt);
\node[above] at (-18pt,0pt) {$-$};
\node[above] at (18pt,0pt) {$+$};
\node[right] at (0pt,-18pt) {$+$};
\end{tikzpicture}
+
\begin{tikzpicture}[scale=1.0, baseline={(current bounding box.center)}]
\filldraw (0pt,0pt) circle (1pt);
\draw[thick] (-20pt,0pt)--(20pt,0pt);
\draw[thick, decorate, decoration={snake, segment length=11pt, amplitude=2pt
}] (0pt,0pt) -- (0pt,-20pt);
\node[above] at (-18pt,0pt) {$-$};
\node[above] at (18pt,0pt) {$-$};
\node[right] at (0pt,-18pt) {$-$};
\end{tikzpicture}
\right)\,.
\label{eqn: vertex}
\end{equation}
In our formalism, only the third term in \eqref{eqn: vertex} will contribute to the rung function for the retarded kernel. Thus, we have
\begin{gather}
\label{Rab}
R_{ab}(t)=\sum_{cc'} \lambda_{abc} \lambda_{bac'}  G_{cc'}^\K (t) \,,
\\
\text{where} \quad G_{cc'}^\K (t)= \Tr \left( \sqrt{\rho} O_c(t)  \sqrt{\rho} O_{c'}(0)\right)\,, \quad \rho = e^{-\beta H}/Z\,.
\end{gather}
To relate $R_{ab}$ to $R^*_{ba}$, we will examine how each factor in \eqref{Rab} is transformed under complex conjugation. In fact, $G_{cc'}^\K (t)$ is real since $O_c$ and $O_{c'}$ are real and the configuration we have chosen is symmetric on the imaginary time circle. As for the coefficients $\lambda_{abc}$, we will use the Hermiticity of the interaction Hamiltonian:
\begin{equation}
H_{\rm int}= \sum_{abc} \lambda_{abc} \psi^\dagger_a \psi_b O_c \,, \quad H_{\rm int} = H^\dagger_{\rm int} \quad \Rightarrow \quad \lambda_{abc} = \lambda_{bac}^*\,.
\end{equation}
It follows that
\begin{equation}
\lambda_{abc} \lambda_{bac'}  = (\lambda_{bac} \lambda_{abc'}  )^*\,.
\end{equation}
Together with the reality of $G^\K$, this implies the desired identity, $R_{ab}(t)=R^*_{ba}(t)$.

\section{Branching time for Brownian SYK}
\label{appendix: brownian}

The Brownian SYK \cite{Saad:2018bqo,Sunderhauf:2019djv} and the regular SYK have the same form for  Hamiltonian
\begin{equation}
H(t) = i ^{\frac{q}{2}} \sum_{j_1<\ldots <j_q} J_{j_1 ... j_q} (t) \psi_{j_1} \ldots \psi_{j_q} \,, 
\end{equation}
but the random ensembles for the couplings are different:
\begin{equation}
\begin{aligned}
\text{regular:} \quad &\bar{ J_{j_1 ... j_q} (t)J_{j'_1 ... j'_q} (t')}= \delta_{j_1j_1'} \ldots \delta_{j_q j_q'} \frac{J^2(q-1)!}{N^{q-1}} \,,  \\
\text{Brownian:} \quad &\bar{ J_{j_1 ... j_q} (t)J_{j'_1 ... j'_q} (t')}= \delta_{j_1j_1'} \ldots \delta_{j_q j_q'} f(t-t')\frac{J(q-1)!}{N^{q-1}} \,. 
\end{aligned}
\end{equation}
Here $f$ is an even narrowly peaked function with integral equal to $1$, so that $f(t)\approx\delta(t)$ for most purposes.

The retarded self-energy $\Sigma^\R$ has the following form for the Brownian SYK (by analogy with the regular SYK):
\begin{equation}
\Sigma^\R(t_1,t_2) = 
\frac{i^{q+2}  J f(t_1-t_2)}{2^{q-1}} \left[\left( G^\R (t_1,t_2) +G^\K(t_1,t_2)\right)^{q-1} +
\left( G^\R (t_1,t_2) -G^\K (t_1,t_2)\right)^{q-1}  
\right]\,.
\end{equation}
Since $f(t_1-t_2)$ enters as an overall factor, we may assume that $t_1-t_2$ is small and use the  UV asymptotics of $G^\R$ and $G^\K$, namely, $G^\R(t) \approx -i \theta(t)$ and $G^\K(t) \approx 0$. Thus,
\begin{equation}
\Sigma^\R(t) = \frac{-i J f(t)}{2^{q-2}} \theta(t)
\approx \frac{-i J}{2^{q-1}}\delta(t)\,.
\end{equation}
This gives the quasiparticle decay rate, 
\begin{equation}
\Gamma := 2i \tilde \Sigma^\R(0) = \frac{J}{2^{q-2}}\,.
\label{eqn: gamma}
\end{equation}
Now, the Keldysh self-energy $\Sigma^\K$ for the Brownian SYK is as follows:
\begin{equation}
\begin{aligned}
\Sigma^\K_{\alpha \beta} (t_1,t_2) = \frac{i^{q+2}  J \delta(t_1-t_2)}{2^{q-1}} &
\left[ \left(G_{\alpha\beta}^\K(t_1,t_2)+  G^\R_{\alpha\beta}(t_1,t_2) -G^\A_{\alpha\beta}(t_1,t_2)  \right)^{q-1} \right. \\
&\left.
+ 
\left(G_{\alpha\beta}^\K(t_1,t_2)-  G^\R_{\alpha\beta}(t_1,t_2) +G^\A_{\alpha\beta}(t_1,t_2)  \right)^{q-1}
  \right] \,.
\end{aligned}
\end{equation}
For the calculation of OTOC and Lyapunov exponent, we are interested in off-diagonal matrix elements, e.g.\ $\alpha=1$,\, $\beta=2$:
\begin{equation}
\Sigma^K_{12} (t) = \frac{i^{q+2}J\delta(t)}{2^{q-2}} \left(G_{12}^\K(t)\right)^{q-1} \,.
\end{equation}
The rung function,
\begin{equation}
R(t) = \frac{\delta \Sigma_{12}^\K}{\delta G_{12}^K} = (q-1) \frac{i^{q+2}J\delta(t)}{2^{q-2}} \left(G_{12}^\K(t)\right)^{q-2} 
\end{equation}
has a delta function factor, and therefore, is an example where the zero-range potential approximation is exact. Finally, we insert the UV form of $G^\K$,
\begin{equation}
G^\K_{12}(0) = 2i \langle \psi \psi \rangle =i
\end{equation}
into the formula for $R(t)$ and get
\begin{equation}
R(t) = (q-1) \frac{J \delta(t)}{2^{q-2}} \,, \quad \tilde R(0) = (q-1) \frac{J}{2^{q-2}} \,.
\end{equation}
Thus, we obtain the Lyapunov exponent and the branching time:
\begin{equation}
\varkappa = \tilde R(0) - \Gamma = (q-2) \frac{J}{2^{q-2}}\,, \quad t_B = \frac{1}{\tilde R(0)} = \frac{2^{q-2}}{(q-1)J} \,.
\end{equation}
The explicit form of the Lyapunov exponent is consistent with the expectation that for $q>2$, the model is chaotic. We have also checked the calculated Lyapunov exponent against the numerical result in \cite{Sunderhauf:2019djv} and found a good agreement.

\section{Branching time for the regular SYK model at large $q$}
\label{appendix: large q}

The large $q$ SYK model was studied in \cite{MS16-remarks}. Starting from the standard SYK Hamiltonian,
\begin{equation}
H= i^{\frac{q}{2}} \sum_{j_1 \ldots < j_q } J_{j_1,\ldots,j_q} \psi_{j_1}  \ldots \psi_{j_q}, \qquad \{\psi_j,\psi_k\}=\delta_{jk},\qquad
\bar{J^2_{j_1,\ldots , j_q}} = \frac{J^2(q-1)!}{N^{q-1}} \,,
\end{equation}
the order of limits is this: first take the $N\to\infty$ limit, and then take $q$ to infinity while keeping $\calJ=\sqrt{2^{1-q}q}\,J$ fixed. In this limit, the two-point function can be computed for any $\calJ$:
\begin{equation}
G(\tau) = -\frac{\sgn(\tau)}{2} \left[ \frac{\cos \frac{\pi v}{2}}{\cos \pi v \left( \frac{1}{2} -\frac{|\tau|}{\beta} \right)}  \right]^{2\Delta}, \qquad \text{where} \quad
\frac{v}{2\cos\frac{\pi v}{2}} =\calJ\,,\quad \beta=2\pi.
\end{equation}
The parameter $v\in (0,1)$ characterizes the coupling strength. 
At strong coupling, $\calJ\gg 1$ and $v\approx 1-\frac{1}{\pi \calJ}$, whereas at weak coupling, $\calJ \ll 1$ and $v\approx 2\calJ$. Roughly speaking, $v$ determines the effective time scale in the system. For example, the quasiparticle decay rate at weak coupling is given by $\Gamma=2v \Delta$ as indicated by the $\tau\to i\infty$ asymptotics of the two-point function. We will also need the rung function,
\begin{equation}
R(t) = \frac{v^2}{2\cosh^2\frac{v t}{2}} \,, \qquad
\tilde R(\omega)=\frac{2\pi\omega}{\sinh(\pi\omega/v)} \,.
\end{equation}

We now compare the exact expressions for the Lyapunov exponent\cite{MS16-remarks} and branching time~\cite{gu2019relation},
\begin{equation}
\varkappa=v \,, \qquad t_B = \frac{3}{2v}\,,
\end{equation}
with the ones derived from the zero-range potential approximation, see section~\ref{section: comments}. Using the fact that $\Gamma$ is small, we get
\begin{equation}
\varkappa_{\text{approx}} = \tilde R(0) -\Gamma \approx 2v \,, \qquad
t_{B,\text{approx}} = \frac{1}{\tilde R(0)}= \frac{1}{2v}\,.
\end{equation}
Note an order $1$ factor discrepancy between the approximate and exact results. This discrepancy is expected since the ``potential term'' $R(t) = \frac{v^2}{2\cosh^2\frac{v t}{2}}$ can not be well approximated by a delta function. 

\bibliography{ref.bib}

\end{document}